\RequirePackage{lineno}
\documentclass[floatfix,twocolumn,prd,showpacs,preprintnumbers,amsmath,nofootinbib,amssymb,noeprint,superscriptaddress]{revtex4-1}

\usepackage[normalem]{ulem}
\newcommand\wave[1]{{\color{blue}{\uwave{#1}}}}

\usepackage{mathtools, cuted}
\usepackage{lipsum}
\usepackage[utf8]{inputenc}
\usepackage{graphicx}
\usepackage{dsfont}
\usepackage{mathrsfs}
\usepackage{bm}
\usepackage{color}
\usepackage[colorlinks,citecolor=blue,urlcolor=blue,linkcolor=black]{hyperref}
\usepackage{braket}
\usepackage{placeins}
\usepackage{multirow}
\usepackage{hyperref}
\usepackage{float}
\usepackage{booktabs}
\usepackage{soul}
\usepackage{float}
\usepackage{comment}

\def\tauf{\tau_{\mathrm{F}}}

\usepackage[nameinlink,capitalize]{cleveref}
\usepackage{diagbox}

\begin{document}

\title{Viscosity of pure-glue QCD from the lattice}

%\date{\today}
\author{Luis Altenkort}
\affiliation{Fakult\"at f\"ur Physik, Universit\"at Bielefeld, D-33615 Bielefeld, Germany }

\author{Alexander M.~Eller}
\affiliation{Institut f\"ur Kernphysik, Technische Universit\"at Darmstadt\\
Schlossgartenstra{\ss}e 2, D-64289 Darmstadt, Germany }

\author{Anthony Francis}
\affiliation{Institute of Physics, National Yang Ming Chiao Tung University, 30010 Hsinchu, Taiwan}

\author{Olaf Kaczmarek}
\affiliation{Fakult\"at f\"ur Physik, Universit\"at Bielefeld, D-33615 Bielefeld, Germany }

\author{Lukas Mazur}
\affiliation{Paderborn Center for Parallel Computing, Paderborn University, D-33098 Paderborn, Germany}

\author{Guy D.~Moore}
\affiliation{Institut f\"ur Kernphysik, Technische Universit\"at Darmstadt\\
Schlossgartenstra{\ss}e 2, D-64289 Darmstadt, Germany }

\author{Hai-Tao Shu}
\email{Corresponding author: hai-tao.shu@ur.de}
\affiliation{Institut f\"ur Theoretische Physik, Universit\"at Regensburg, D-93040 Regensburg, Germany}

\begin{abstract}

We calculate shear viscosity and bulk viscosity in $SU(3)$ gauge theory on the lattice at $1.5 \,T_c$.
The viscosities are extracted via a Kubo formula from the reconstructed spectral function which we determine from the Euclidean time dependence of the corresponding channel of the energy-momentum tensor correlators.
We obtain unprecedented precision for the correlators by applying gradient-flow and blocking methods.
The correlators are extrapolated to the continuum and then to zero-flow time.
To extract the viscosities we fit theoretically inspired models to the lattice data and cross-check the fit results using the Backus-Gilbert method.
The final estimates for shear and bulk viscosity are $\eta/s = 0.15-0.48$ and $\zeta/s = 0.017-0.059$.

\end{abstract}

\maketitle

\section{Introduction}
\label{sec:intro}

The shear viscosity $\eta$ and bulk viscosity $\zeta$ of the hot quark-gluon plasma characterize the dissipation which occurs due to nonuniform flow, such as occurs in heavy ion collisions.
They have been a topic of intense study for the last two decades.
Experimental results \cite{STAR:2000ekf, PHENIX:2003qra, ALICE:2011ab, ATLAS:2014ndd, ALICE:2016kpq} suggest a small shear viscosity;
indeed, based on the determined values of elliptic and higher-order flow as functions of momentum and impact parameter, the best extractions of the shear viscosity are in the range
$1/(4\pi)<\eta/s<2/(4\pi)$ \cite{JETSCAPE:2020mzn}.
%\todoolaf{use a consistent way for quoting $\eta$, either $\eta/s$ or $(4\pi)\eta/s$.}
This is close to the claimed lower bound on $\eta/s$ obtained from $\mathcal{N}=4$ supersymmetric Yang-Mills theory at strong coupling,
which predicts $\eta/s = 1/(4\pi)$ \cite{Policastro:2001yc}.
While leading-order weak-coupling calculations \cite{Arnold:2000dr,Arnold:2003zc}, extrapolated to the physical coupling strength, suggest a larger shear viscosity $\eta/s \sim 0.5$--$1$, the next-to-leading correction to this result at a physically interesting coupling and temperature reduces the tension, implying $\eta/s \sim 0.2$ \cite{Ghiglieri:2018dib}.
The size of this difference implies that the perturbative series shows poor convergence.
As for the bulk viscosity, its extraction from experiments shows that it is nonzero but somewhat smaller than the shear viscosity at temperatures of order 200 MeV \cite{JETSCAPE:2020mzn}.
At higher temperatures we have a leading-order perturbative calculation
\cite{Arnold:2006fz} which shows that, for $0.06<\alpha_s<0.3$, $\zeta/s\sim 0.02\alpha_s^2$.
That is, as the theory becomes more conformal at higher temperatures, the bulk viscosity is expected to become small, but it can nevertheless play a role at lower temperatures where QCD behaves strongly nonconformally.

We want a first-principles theoretical determinations of shear and bulk viscosity, to accompany the values extracted from experiment.
The temperatures achieved in real-world heavy ion collisions are in a range where perturbation theory does not appear to be applicable, and so truly nonperturbative methods are needed.
Our best first-principles nonperturbative tool is lattice gauge theory, which we will pursue in this work.
Like previous literature, we will work within pure $SU(3)$ gauge theory, but one focus of our work is to develop tools which will be straightforward to extend to the theory with dynamical quarks.

The pioneering works \cite{Nakamura:2004sy, Meyer:2007ic,Meyer:2007dy} established the general approach for investigating shear viscosity via unequal Euclidean-time, zero space-momentum energy-momentum tensor (EMT) correlation functions.
More recent studies~\cite{Astrakhantsev:2017nrs,Astrakhantsev:2018oue} have extended this work to consider a range of temperatures.
However, these works used rather coarse and small lattices, meaning that  cutoff effects may be severe.
Recently, a lattice calculation using the gradient flow method was conducted on a $64^3\times 16$ lattice \cite{Itou:2020azb}.
In that work, the shear viscosity is extracted at finite flow time, making the results difficult to interpret \cite{Altenkort:2020fgs}.

The standard way to investigate transport coefficients on the lattice is through Kubo formulas, which relate these coefficients to spectral functions, which in turn are related to Euclidean correlators through analytic continuation.
The biggest challenge is that the energy-momentum tensor correlators, from which the viscosities are extracted, are extremely noisy, such that a noise-reduction technique must be employed to obtain the necessary precision.
In Refs. \cite{Meyer:2007dy, Astrakhantsev:2018oue} the multilevel algorithm~\cite{Luscher:2001up} was used; in this work we instead make use of the gradient-flow method ~\cite{Luscher:2010iy,Luscher:2013cpa,Luscher:2010we,Narayanan:2006rf} and the blocking method~\cite{Altenkort:2021jbk} which we proposed recently.
In comparison to multilevel algorithms, the gradient flow approach has the advantages that it is straightforward to apply to the full theory with dynamical quarks, and it helps with the problem of operator renormalization.
This paves the way for a future study in full QCD.
The signal is improved further via the blocking method, up to a factor of 7 without additional computation cost, as we demonstrate in ~\cite{Altenkort:2021jbk}.
With these two methods we are able to achieve high precision for the desired correlators. 

Our lattice setup consists of five large and fine lattices, of which the coarsest one ($64^3\times 16$) is already as large as the finest lattice used in previous literature.
The largest and finest lattice in our study is of size $144^3\times 36$ at $\beta=7.544$ ($a=0.0117\mathrm{fm}$).
With our setup, including such a fine lattice, the continuum extrapolation is well-behaved and, thanks to the large temporal extents of the underlying lattices, the results of the spectral reconstruction will be more reliable.

In the following we will start with the definition of the EMT under gradient flow and explain how shear and bulk viscosity can be obtained from the EMT correlators. 
In Sec. III we give the lattice setup used in this study. Sec. IV is devoted to the nonperturbative renormalization of the EMT correlators. After a short illustration to the temperature-correction and tree-level improvement in Sec. V we continue with the discussions of continuum extrapolation and flow-time extrapolation in Sec. VI. In Sec. VII we focus on the extraction of viscosities via spectral analysis and provide our estimates for the viscosities. The conclusion is given in Sec. VIII.

%\todoguy{Summary of content of the paper here??}

\section{Transport, energy-momentum tensor, and gradient flow}
\label{sec:gradflow}

The fundamental object of our study is the energy-momentum tensor $T_{\mu\nu}$, defined as the Noether current of 4-translation symmetry (or equivalently as the variation of the action with respect to the spacetime metric).
Shear viscosity is the response of $T_{ij}$ to shear flow, under which the traceless part of $\partial_i v_j$ is nonzero.
Shear flow also couples to the energy-momentum tensor, so the Kubo relation describing the shear viscosity involves a correlation function of two traceless energy-momentum tensors,
\begin{align}
\label{kubo1}
\eta(T) & =\lim_{\omega\rightarrow  0}
\frac{\rho_{\rm{shear}}(\omega,T)}{\omega},
\\ \nonumber
\rho_{\rm{shear}}(\omega,T) & = \frac{1}{10} \int \mathrm{d}^3 x \, \mathrm{d}t \,
e^{i\omega t} \left\langle \left[ \pi_{ij}(x,t) \,,\, \pi_{ij}(0,0) \right]
\right\rangle \,,
\\ \noindent\pi_{ij} & = T_{ij} - \frac{1}{3} \delta_{ij} T_{kk} \,.
\end{align}
Similarly, bulk viscosity is the response of the trace of the energy-momentum tensor to a divergent fluid flow, which also couples to the trace of the energy-momentum tensor,
\begin{align}
\label{kubo2}
    \zeta(T) & = \frac{1}{9} \lim_{\omega \to 0} 
\frac{\rho_{\rm{bulk}}(\omega,T)}{\omega} ,
\\
\rho_{\rm{bulk}} & = \int \mathrm{d}^3 x \, \mathrm{d}t \, e^{i\omega t}
\left\langle \left[ T_{\mu\mu}(x,t) \,,\, T_{\nu\nu}(0,0) \right] \right\rangle .
\end{align}

%\todoolaf{Does the bulk mode also contain temporal components? Or should it be $T_{ii} T_{jj}$ here?}\todohaitao{yes. as said in previous sentence it's the correlation of the trace, so it contains the temporal contribution.}
%\todoguy{The $T_{00}$ correlators are time-independent so they strictly give a frequency delta-function which doesn't appear in the $\omega \to 0$ limit.  So you can add them or leave them out.  It's better to put them in because it makes the desired operator simpler.}
Our approach will be to use analyticity to relate these spectral functions to the Euclidean, time-dependent correlation (still at zero momentum or equivalently with $\int \mathrm{d}^3 x$),
\begin{align}  
\label{Gthetatau}
G(\tau)=\int_0^{\infty}\frac{\mathrm{d}\omega}{\pi}\frac{\cosh[\omega(1/2T-\tau)]}{\sinh(\omega/2T)}\rho(\omega,T).
\end{align} 
This expression can in principle be inverted to determine the spectral function, a task we will return to in Sec. \ref{sec:spectrum}.
Here $G(\tau)$ is the Euclidean function associated to the respective spectral function, that is,
\begin{align} 
\begin{split}
&G_{\rm{shear}}(\tau)=\frac{1}{10} \int \mathrm{d}^3x\ \left\langle \pi_{ij}(0,\vec{0}) \: \pi_{ij}(\tau,\vec x)
 \right\rangle,\\
&G_{\rm{bulk}}(\tau)=\int \mathrm{d}^3x\ \left\langle T_{\mu\mu}(0,\vec{0}) \: T_{\mu\mu}(\tau,\vec{x})\right\rangle.
\label{eq:Gshearbulk}
\end{split}
\end{align}
Our main task will be evaluating the continuum limit of these correlation functions precisely.

There are two principle challenges when treating energy-momentum tensor correlations on the lattice:  the correlations are very noisy, and because of the lack of continuous translation symmetry on the lattice, there is no obvious choice for the energy-momentum tensor operator.
In particular, different components of $\pi_{ij}$ renormalize differently, which presents a challenge.
Both problems are ameliorated if we utilize gradient flow to generate our energy-momentum operators.
Gradient flow is defined as the iterative replacement of the gauge fields  $A_\mu(x)$ with fields containing less UV fluctuations, $B_\mu(x,\tauf)$, through the definitions \cite{Luscher:2010iy}
\begin{align}
\label{flow_equation}
B_\nu(x,\tauf=0) & = A_\nu(x) \,,
\nonumber \\
\dot{B}_{\mu} & = D_{\nu}G_{\nu\mu},
\nonumber \\
G_{\mu\nu} & = \partial_{\mu}B_{\nu}-\partial_{\nu}B_{\mu}+[B_{\mu},B_{\nu}],
\nonumber \\
D_{\mu} & = \partial_{\mu}+[B_{\mu},\cdot].
\end{align}
That is, at $\tauf=0$ the flowed field is the nonflowed field, but the field then evolves under a covariant heat equation which iteratively removes the most UV fluctuations of the field.
Using the flowed field to construct operators such as the energy-momentum tensor leads to operators with well behaved renormalization properties and improved rotational invariance. In terms of the gradient-flowed field, we define the gradient-flowed squared field strength operator and the traceless tensor operator as
\begin{equation}
\label{E_U}
\begin{split}
E(\tauf,x)&=\frac{1}{4}F^a_{\rho\sigma}(x,\tauf)F^a_{\rho\sigma}(x,\tauf),\\
U_{\mu\nu}(x,\tauf)&=F^a_{\mu\rho}(x,\tauf)F^a_{\nu\rho}(x,\tauf)-\delta_{\mu\nu}E(\tauf,x).
\end{split}
\end{equation}
The energy-momentum tensor can then be written in  terms of these two operators and two not yet known coefficients as
~\cite{Suzuki:2013gza}
\begin{equation}
\label{EMT_flow}
T_{\mu\nu}(\tauf,x)=c_1(\tauf) U_{\mu\nu}(\tauf,x)+4c_2(\tauf)\delta_{\mu\nu}E(\tauf,x).
\end{equation}
Here $c_1$, $c_2$ are the coefficients on the traceless and pure-trace parts of the tensor, respectively.
Arguably one should perform a vacuum subtraction from $E(\tauf,x)$, but in practice we always compute connected correlation functions, which implements such a subtraction automatically.

There are two approaches to determining the coefficients $c_1,c_2(\tauf)$.
Suzuki has determined them up to 2-loop and 3-loop order in the $\overline{\mathrm{MS}}$-scheme~\cite{Suzuki:2021tlr}:
\begin{align}
\label{coeff_UE1}
c_1^{\text{(N${}^2$LO)}}(\tauf) & =
\frac{1}{g^2(\mu)} \sum_{n=0}^{2} k_1^{(n)}(L(\mu,\tauf)) \Big{[} \frac{g^2(\mu)}{(4 \pi)^2 } \Big{]}^n ,\\
c_2^{\text{(N${}^3$LO)}}(\tauf) & =
\frac{1}{g^2(\mu)} \sum_{n=1}^{4} k_2^{(n)}(L(\mu,\tauf)) \Big{[} \frac{g^2(\mu)}{(4 \pi)^2 } \Big{]}^n,
\label{coeff_UE2}
\end{align}
where the coefficients $k_1^{(n)}$, $k_2^{(n)}$ can be found in \cite{Harlander:2018zpi,Iritani:2018idk}. 
Here $L(\mu,\tauf)\equiv \log(2\mu^2e^{\gamma_E}\tauf)$ and the running coupling can be evaluated in the $\overline{\mathrm{MS}}$-scheme at scale $\mu=1/ \sqrt{8\tauf}$~\cite{Harlander:2016vzb}. 
The series for $c_2$ begins with a constant and is known to one higher order than for $c_1$; therefore it suffers very little coupling and renormalization-point uncertainty, and is more accurate than any numerics-based nonperturbative estimate which we could develop.
Therefore, we use the series expansion for $c_2$.
The error in this series expansion is negligible, below 0.1\%.
This will be swamped by statistical errors in our correlation functions and will play no role in our error analysis.

In contrast, since $c_1$ depends on the coupling at leading order, the use of a series expansion is significantly less reliable.
Instead, we will perform a nonperturbative renormalization on the lattice in 
Sec.\ref{sec_renormalization}, based on ideas developed by Giusti and Pepe \cite{Giusti:2015daa}.

According to small-flow time expansion~\cite{Luscher:2011bx}, any composite operator at finite flow time can be expressed as superposition of renormalized operators with finite, flow-dependent coefficients~\cite{DelDebbio:2013zaa}.
That is, one can expand our stress tensor operator in an operator product expansion, where the first term is the desired stress-tensor and higher terms represent various higher-dimension operators with coefficients containing positive powers of $\tauf$.
Therefore, one expects that the correlation functions we evaluate, at separation $\tau$, correspond to the correct correlation functions, plus corrections which appear as a series expansion in $(\tauf/\tau^2)$.
Determining the desired correlation function therefore requires an extrapolation to $\tauf \to 0$ to eliminate the effects of these high-dimension contaminants.
Only some finite range of $\tauf$ values will actually be useful in this extrapolation; larger values of $\tauf$, such that $(\tauf/\tau^2)$ is not small, will be outside of the range where an extrapolation is possible.
Solving Eq.~(\ref{flow_equation}) perturbatively suggests that the flow smears the gauge field with a radius 
$r \simeq \sqrt{8\tauf}$~\cite{Luscher:2010iy}. 
In general this radius should be larger than one lattice spacing to suppress the lattice effects and noise, and at the same time smaller than half the lattice extent so that the flow radius does not interact with the lattice periodicity.
For a specific operator there can be further constraints on the flow radius.
How much flow can be applied and what Ansatz should be used for the $\tauf\rightarrow 0$ extrapolation will be discussed in a later section.

\section{Lattice Setup}
\label{sec:latt}

Our lattice calculations are carried out in $SU(3)$ Yang-Mills theory in four-dimensional spacetime with periodic boundary conditions for all directions. We summarize the settings in Table~\ref{tab:lattice_setup}. The gauge configurations are generated using the standard Wilson gauge action on five large, fine, isotropic lattices. On each lattice we generate 10,000 configurations. To ensure the gauge fields are fully thermalized the first 4,000 sweeps (each consists of one heat bath and four over-relaxation steps) are discarded. In the sampling procedure the configurations are stored after every 500 sweeps. This removes the autocorrelations in observables as we have confirmed. All the lattices are set to the same temperature $\sim 1.5T_c$ by tuning the $\beta$ value. The scale is set via the Sommer parameter $r_0$~\cite{Sommer:1993ce} with state-of-the-art value $r_0 T_c = 0.7457$~\cite{Francis:2015lha}. The  parametrization form needed in scale setting is taken from~\cite{Francis:2015lha} with updated coefficients from~\cite{Burnier:2017bod}.
\begin{table}[htb]       
    \centering
    \begin{tabular}{ccrccccc}                            
    \hline \hline
    $a$ (fm) & $a^{-1}$ (GeV) & $N_{\sigma}$ & $n_{\sigma}$ & $N_{\tau}$ & $\beta$ & $T/T_{c}$ & \#Configuration\tabularnewline
    \hline
    0.0262 & 7.534 & 64 & 4 & 16  & 6.8736 &  1.5104  & 10000 \tabularnewline
    0.0215 & 9.187 & 80 & 4 & 20  & 7.0350 &  1.4734  & 10000 \tabularnewline
    0.0178 & 11.11 & 96 & 4 & 24  & 7.1920 &  1.4848  & 10000 \tabularnewline
    0.0140 & 14.14 & 120 & 6 & 30  & 7.3940 &  1.5118  & 10000 \tabularnewline
    0.0117 & 16.88 & 144 & 8 & 36  & 7.5440 &  1.5042  & 10000 \tabularnewline
    \hline \hline
    \end{tabular}
    \caption{ $\beta$ values, lattice spacings, lattice sizes, blocking bin size $n_{\sigma}$ and number of configurations in this study.
    \label{tab:lattice_setup}}
\end{table}

%\begin{figure}[hbt] 
%\centerline{\includegraphics[width=0.5\textwidth]{./ZU_norm.pdf}
%}
%\caption{The normalization constants $c_1(\tauf)$ calculated non-perturbatively using $\langle \epsilon+P \rangle$ and perturbatively up to 2-loop order\cite{Suzuki:2021tlr} on different lattice.
%}
%\label{fig-ZU}
%\end{figure}

We use the clover definition of the energy-momentum tensor appearing in Eq.~(\ref{E_U}).
%Note that, on the lattice, one should actually distinguish between the 6 nondiagonal and the 3 traceless but diagonal stress tensor elements, which renormalize with distinct coefficients \cite{Caracciolo:1989pt,Caracciolo:1991cp,Giusti:2015daa}.
%In practice we compute $c_1$ for the diagonal components and use it for both types of stress tensor operator.
%This is justified because the ratio of the two renormalization constants approaches 1 in the continuum limit at fixed flow depth with $\mathcal{O}(a^2)$ errors; furthermore, a recent lattice determination of this ratio in the flow range of interest here shows that it is already consistent with 1 at the smallest flow depth which we use%
The gradient flow is a Symanzik improved version~\cite{Ramos:2015baa}.
We measure the EMT correlators at 140 discrete flow times in the range 
%\todoluis{Maybe short comment on the (adaptive) step sizes?} \todohaitao{would that be too technical?} 
$\sqrt{8\tauf}T\in \lbrace 0.004, \dots, 0.375\rbrace$ using an adaptive step-size method.
In this method the step size is updated after each integration step such that the error in the integration does not exceed a certain tolerance \cite{Fritzsch:2013je}.
The bin size used in the blocking method is given as $n_{\sigma}$ in Table~\ref{tab:lattice_setup}.

\section{Renormalization}
\label{sec_renormalization}

In this section we describe how we determine the renormalization constants appearing in Eq.~(\ref{EMT_flow}).
We determine the constant $c_1$ using a method inspired by the work of Giusti and Pepe \cite{Giusti:2015daa}.
Namely, we observe that the enthalpy density is given by
\begin{equation}
\label{entropy}
    %\todoluis{should we include a sum over $x$ here?}
    \langle \epsilon+P \rangle_{\tauf}= c_1(\tauf) \left\langle   \frac{1}{3}U_{ii}(\tauf)- U_{00}(\tauf)\right\rangle ,
\end{equation}
where ``0'' denotes the time direction.
Since $\epsilon + P$ has been measured at the sub-percent level
\cite{Giusti:2016iqr}, we can determine $c_1$ through the ratio
$c_1(\tauf)=\langle \epsilon+P \rangle_{\tauf}/\langle  \frac{1}{3}U_{ii}(\tauf)- U_{00}(\tauf)\rangle$.
We will explain below why this also determines the coefficients for the off-diagonal components of the stress tensor, to sufficient precision for this work.

\begin{figure}[tbh]
\centerline{\includegraphics[width=0.5\textwidth]{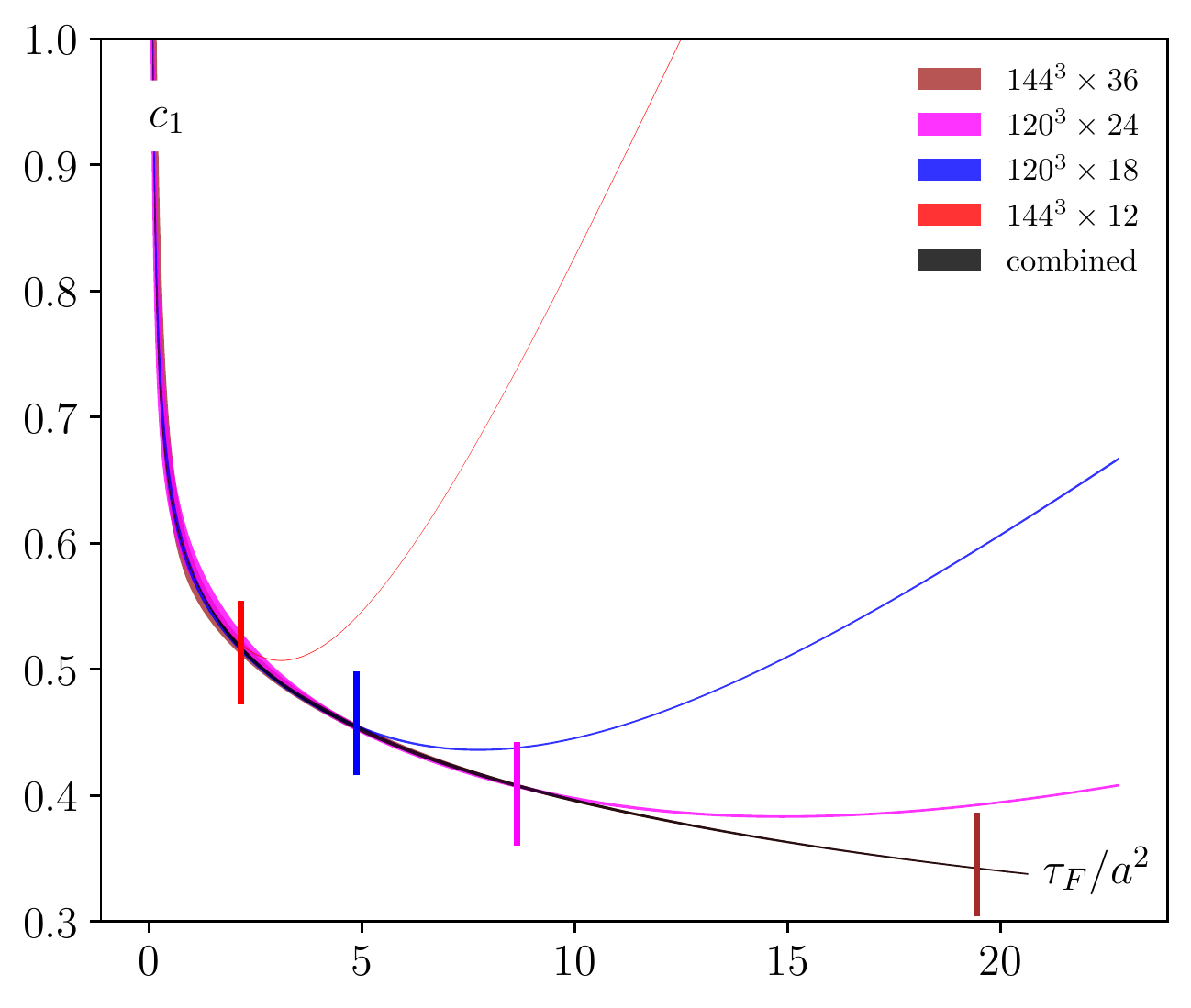}}
\caption{$c_1$ measured at several higher temperatures at $\beta=7.544$ and their weighted average.  The vertical bars indicate the flow depth where each $N_\tau$ choice is expected to become unreliable. }
\label{fig_ZU_highT}
\end{figure}

Unfortunately the enthalpy density is proportional to $T^4$ and therefore to $N_{\tau}^{-4}$, which leads to a poor signal-to-noise ratio for the finest lattices.
We overcome this limitation by measuring $\epsilon+P$ at a range of $N_\tau$ values listed in Table \ref{tab:highT}, not just the ones given in Table~\ref{tab:lattice_setup}.
This is possible because the renormalization constant $c_1$ depends on the lattice spacing but not on the temperature.
However, after enough gradient flow, the gradient flow radius starts to interact with the periodicity radius and the result becomes contaminated and unreliable.
A leading-order perturbative estimate of this effect is that \cite{Eller:2018yje}
\begin{align}
\label{Ellerparam}
\frac{\langle  \frac{1}{3}U_{ii}- U_{00}\rangle_{\mathrm{flowed}}}{(\frac{1}{3}U_{ii}- U_{00})_{\mathrm{true}}} 
& =  1 - \frac{180}{\pi^4} e^{-1/x}
\left( 1 + \frac{1}{x} + \frac{1}{2x^2}\right),
\nonumber \\
\mbox{with} \hspace{2em} x & = 8\tauf T^2 \,.
\end{align}

\begin{figure*}[tbh] 
\centerline{\includegraphics[width=0.5\textwidth]{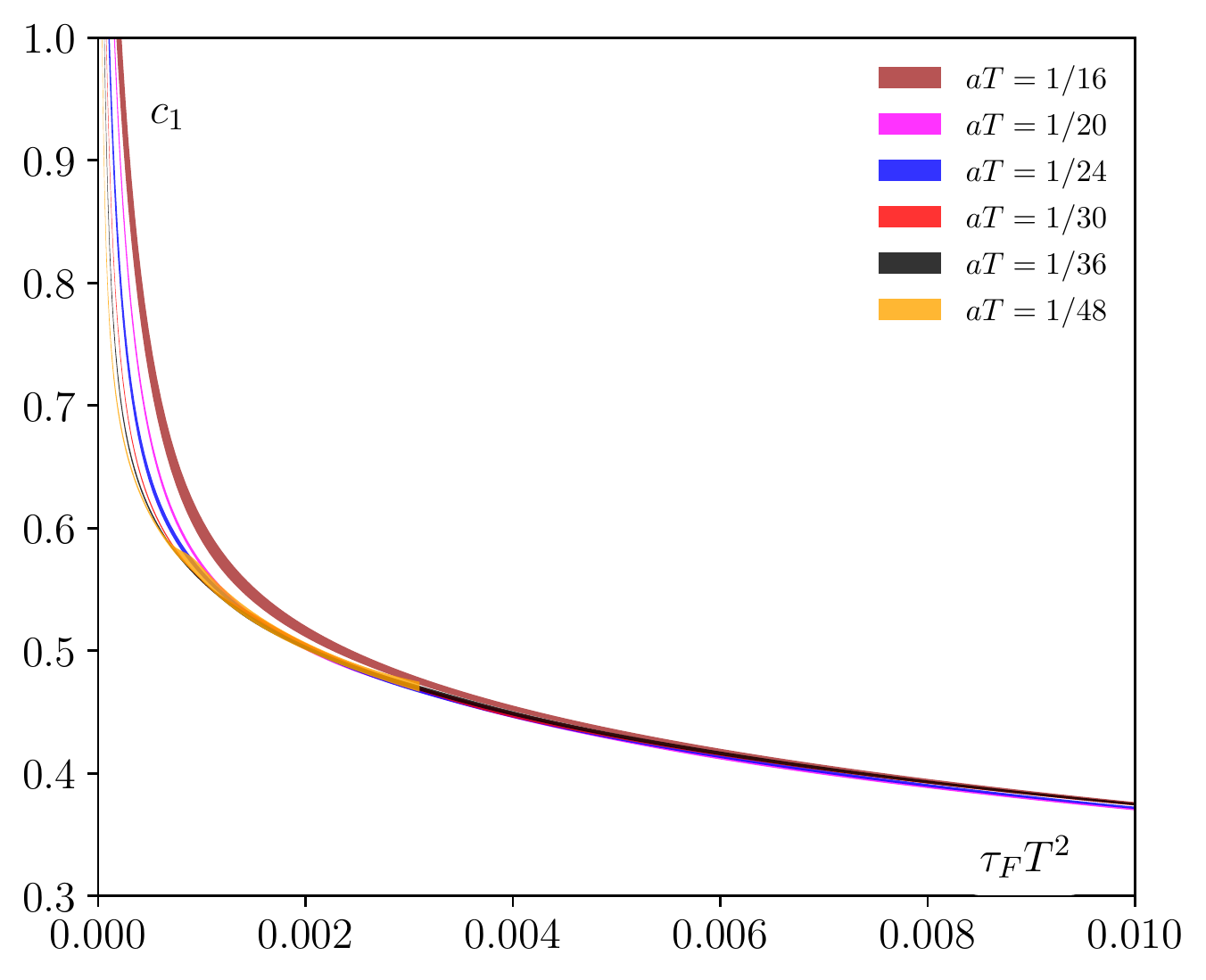}\includegraphics[width=0.5\textwidth]{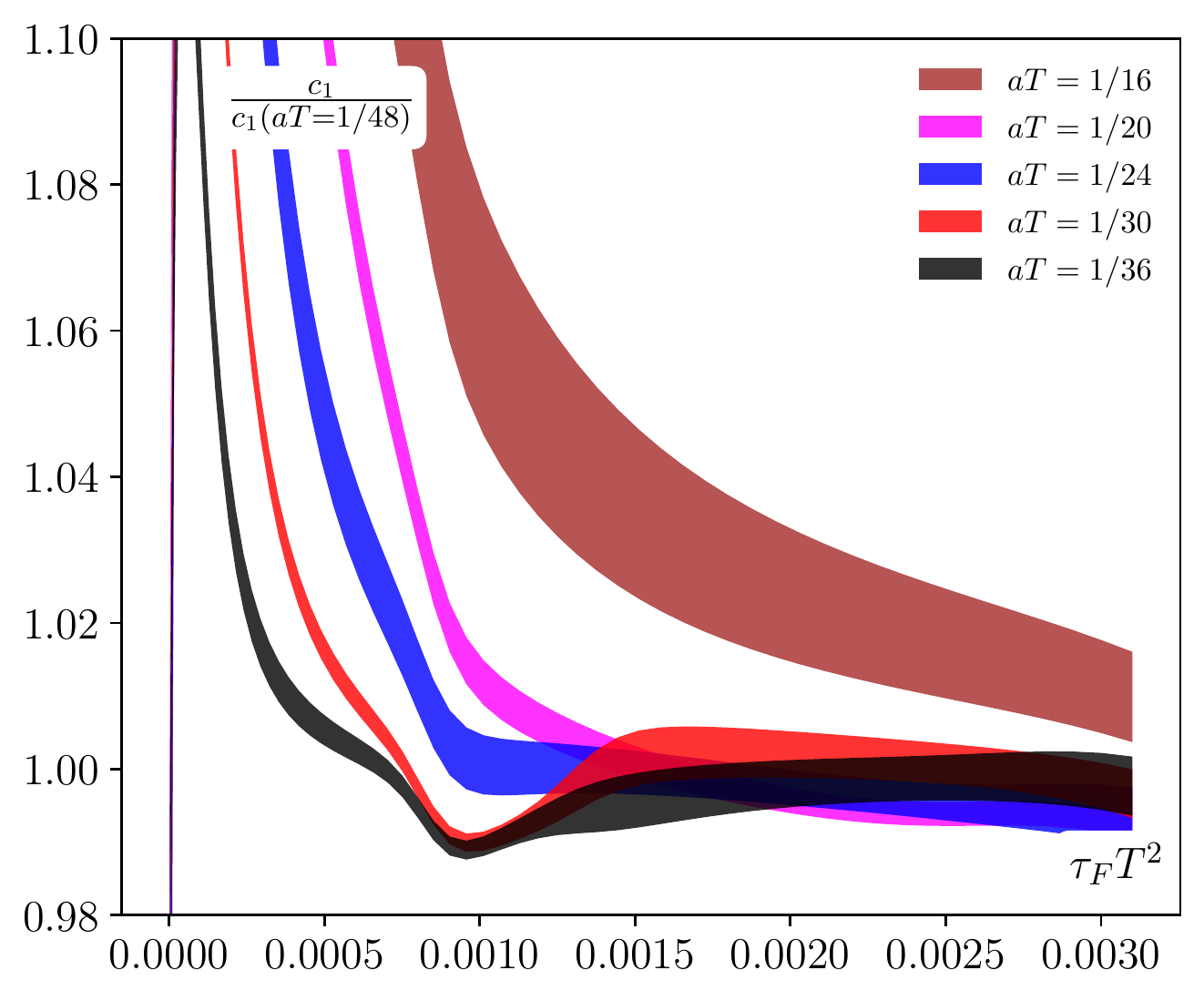}}
\caption{Left: combined $c_1$ at different lattice spacings. Right: the ratio of $c_1/c_1(\beta=7.793)$. The error in the estimation of $c_1(\beta=7.793)$ is not included in the ratio, because $c_1(\beta=7.793)$ is only used as a normalization and its error is irrelevant to the rest analysis. The temperature $T$ in the legends $aT$ and $\tauf T^2$ has been fixed to 1.5$T_c$.
}
\label{final_c1}
\end{figure*}

We illustrate the method, and the effect of the different $N_\tau$ choices, in Fig.~\ref{fig_ZU_highT}, which shows $c_1$ for our finest lattice at different temperatures.
It can be seen that, at very small flow times $c_1$, measurements from different temperatures agree with each other, with smaller statistical errors for the smaller $N_\tau$ values.
With increasing flow time, the higher-temperature $c_1$ values start to deviate from the lower ones.
The point where Eq.~(\ref{Ellerparam}) implies a 1\% correction is marked for each $N_\tau$ value by a vertical bar, and it corresponds well with the flow time value where a given lattice starts to deviate clearly from the larger-$N_\tau$ lattices.

Our final estimate for $c_1$ will be based on a weighted average of the value determined from each $N_\tau$ value we explored.
The weight is determined as
$1/(\sigma_{\textrm{stat}}+\sigma_{\textrm{syst}})^2$ where $\sigma_{\textrm{stat}}$ is the statistical uncertainty from the lattice data and $\sigma_{\textrm{syst}}$ is the systematic
shift as determined from Eq.~(\ref{Ellerparam}).
The averaged $c_1$ is the black curve labeled ``combined" in Fig.~\ref{fig_ZU_highT}.

\begin{table}[ht]      
\centering
\begin{tabular}{ccccc}
\hline \hline
$\beta$ & $a$[fm]($a^{-1}$[GeV]) & $N^h_\tau$ & $N^h_{\sigma}$ & $\#$Configuration\\ \hline
6.8736 &  0.0262 (7.534) & 12 & 64 & 1000\\ \hline
\multirow{2}{*}{7.0350} &\multirow{2}{*}{0.0215 (9.187)} & 10 & 80 & 1000\\
    & & 14 & 80 & 1000\\ \hline
\multirow{2}{*}{7.1920} & \multirow{2}{*}{0.0178 (11.11)} & 12 & 96 & 1000\\
    & & 18 & 96 & 1000\\ \hline
\multirow{2}{*}{7.3940} &  \multirow{2}{*}{0.0140 (14.14)} & 10 & 120 & 1000\\
    & & 16 &120 & 1000\\ \hline 
\multirow{3}{*}{7.5440} & \multirow{3}{*}{0.0117 (16.88)} & 12 & 140 & 1000\\
    & & 18 &120 & 1000\\
    & & 24 & 120 & 1000\\ \hline 
\multirow{3}{*}{7.7930} & \multirow{3}{*}{0.0087 (22.78)} & 12 & 144 & 500\\
    & & 24 &144 & 500\\
    & & 48 & 192 & 700\\ \hline \hline
\end{tabular}
\caption{The lattices with smaller temporal extents for the determination of $c_1$.} 
\label{tab:highT}
\end{table}

We repeat this procedure for the other lattice spacings and summarize the final $c_1$ in Fig.~\ref{final_c1}.
The statistical error in $c_1$ is small, ranging from $1.1\%$ at the smallest flow time we use to $0.27\%$ at the largest flow time we use.
A table presenting the statistical uncertainties of $c_1$ at each lattice spacing for a range of flow times is provided in Appendix \ref{sec:error_renorm}.

Let us now focus on the small flow-time region, to establish how much flow time is enough to eliminate lattice spacing effects.
We have added one more, still finer lattice ($\beta=7.793$, with $N_\tau=48$ when $T/T_c = 1.5$) so that we can compare to a still more continuumlike case.
We can see that lattice cutoff effects are suppressed at large flow times but at small flow times they are noticeable. 
To see down to what flow time the $c_1$ is free of lattice cutoff effects, we plot the ratio $c_1/c_1(\beta=7.793)$ in the right panel.
In order to see more clearly how the different lattice spacings differ from each other, we have plotted error bars based only on the statistical errors in the coarser lattices -- that is, statistical errors in the $\beta=7.793$ lattice are treated as a common systematic error in the right plot.
The figure shows that the lattices give compatible $c_1$ values as long as the flow time is large enough; but each lattice starts to deviate at a flow time such that $\tauf/a^2$ becomes order one.
Specifically, in every case the deviation from continuum behavior reaches 2\% when $\tauf \simeq 0.4 a^2$.
The deviation rapidly becomes more severe below this point.
This deviation from continuum behavior indicates that the applied gradient flow is not sufficient to supply a continuumlike, well-renormalized stress-tensor operator.
Since the statistical precision of our EMT correlator data is typically around 2\% and since we want to keep systematic effects smaller than this,
we will impose the condition
$\tauf \geq 0.4 a^2$ when we perform the double extrapolation of shear correlators in the next section.

\begin{table}[htb]       
    \centering
    \begin{tabular}{ccrccccc}                            
    \hline \hline
    $a$ (fm) & $a^{-1}$ (GeV) & $N_{\sigma}$ & $N_{\tau}$ & $\beta$ & $T/T_{c}$ & \#Configuration\tabularnewline
    \hline
    0.0262 & 7.534 & 64 &  64  & 6.8736 &  0.3776  & 1000 \tabularnewline
    0.0215 & 9.187 & 80 &  80  & 7.0350 &  0.3684  & 1000 \tabularnewline
    0.0178 & 11.11 & 96 & 96  & 7.1920 &  0.3712  & 1000 \tabularnewline
    0.0140 & 14.14 & 96 & 120  & 7.3940 &  0.3780  & 1000 \tabularnewline
    0.0117 & 16.88 & 96 & 144  & 7.5440 &  0.3761  & 1000 \tabularnewline
    \hline \hline
    \end{tabular}
    \caption{ The lattices at $T<T_c$ for the study of $c_2$.}
    \label{tab_zeroT}
\end{table}

\begin{figure}[htb] 
\centerline{\includegraphics[width=0.5\textwidth]{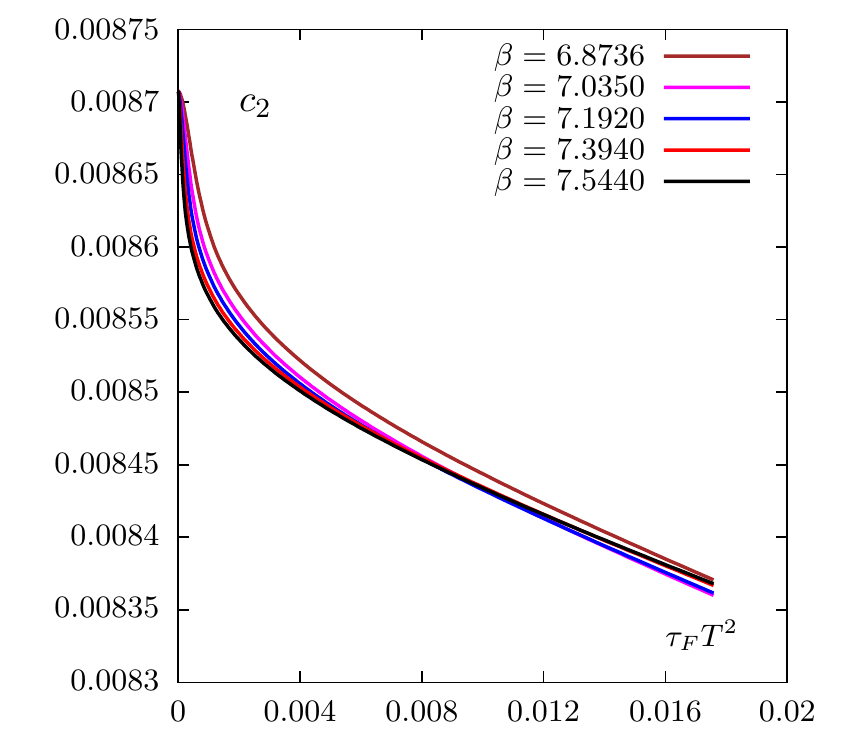}
}
\caption{$c_2$ measured at $T<T_c$ at different lattice spacings.
}
\label{fig_c2}
\end{figure}

Now we calculate $c_2$. According to Eq.~(\ref{coeff_UE2}), the running coupling in the $\overline{\mathrm{MS}}$ scheme is needed. For that we first calculate the coupling in the gradient-flow scheme and then convert it to the  $\overline{\mathrm{MS}}$ scheme. In the gradient-flow scheme the running coupling can be calculated as \cite{Fodor:2012td,Hasenfratz:2019hpg}
\begin{align}
\label{g2flow}
    g_{\mathrm{flow}}^2=\frac{128\pi^2}{3(N_c^2-1)}\frac{1}{1+\delta(\tauf)}\langle \tauf^2 E\rangle,
\end{align}
where $N_c=3$ and $E$ is the energy density defined in Eq.~(\ref{E_U}). $\delta(\tauf)$ can be found in \cite{Fodor:2012td,Hasenfratz:2019hpg} as well. Note that the energy density should be measured at zero temperature. On the lattice we take large temporal extents to suppress the thermal effects.  %\todoluis{how do we know?} which should be no big change from zero temperature. 
The lattices used to study this quantity are given in Table~\ref{tab_zeroT}.
Because of high computation costs the two finest lattices have smaller-aspect ratios.
However, based on the three coarse lattices, we have seen that finite volume effects are small compared to the statistical error of the correlators.

After obtaining $\tauf^2E$ in the gradient flow scheme, we can relate it to the one in the $\overline{\mathrm{MS}}$ scheme \cite{Harlander:2016vzb}. This requires solving a cubic equation, whose solution gives the running coupling in the $\overline{\mathrm{MS}}$ scheme. Inserting in Eq.~(\ref{coeff_UE2}), we get the final $c_2$ shown in Fig.~\ref{fig_c2}. The errors are not visible as they are tiny and in every case much smaller than 1\%. We can see that unlike $c_1$, the difference of $c_2$ among different lattice spacings is always small. The ratio $c_2/c_2(\beta=7.544)$ is always smaller than 1\% at all flow times, suggesting that the cutoff effects can be ignored for $c_2$.

\section{large separations and noise reduction}

Evaluating Eq.~(\ref{eq:Gshearbulk}) involves computing a correlator with an integral over all values of the spatial separation.
To improve signal-to-noise ratio, in practice one evaluates $\int \mathrm{d}^3 x \mathrm{d}^3 y \, \mathrm{d}t \langle T(x,\tau+t) T(y,t) \rangle$,
that is, one performs an integral over the coordinates of each operator.
The correlation function is dominated by small values of coordinate difference $|x-y|$.
However, the fluctuations in the correlator, and therefore the noise, are approximately separation independent.
Therefore, the inclusion of large separations makes the evaluation noisy without contributing meaningfully to the signal.

In Ref.\cite{Altenkort:2021jbk} we proposed a way to reduce these noise contributions.
The operator of interest ($T_{\mu\mu}$ or a component of $\pi_{ij}$) is first summed over small volumes called blocks, on a single $\tau$ sheet but with a cubic space extent given in Table~\ref{tab:lattice_setup}.
We evaluate all block-to-block correlators and then average all correlators which have the same temporal and block-center spatial separation.
Finally, we examine how the correlation function varies with the space separation between blocks, replacing the large-separation, small-signal values with an asymptotic fit as described in \cite{Altenkort:2021jbk}.

\begin{figure}[tb]
%\medskip
\includegraphics[width=0.482\textwidth]{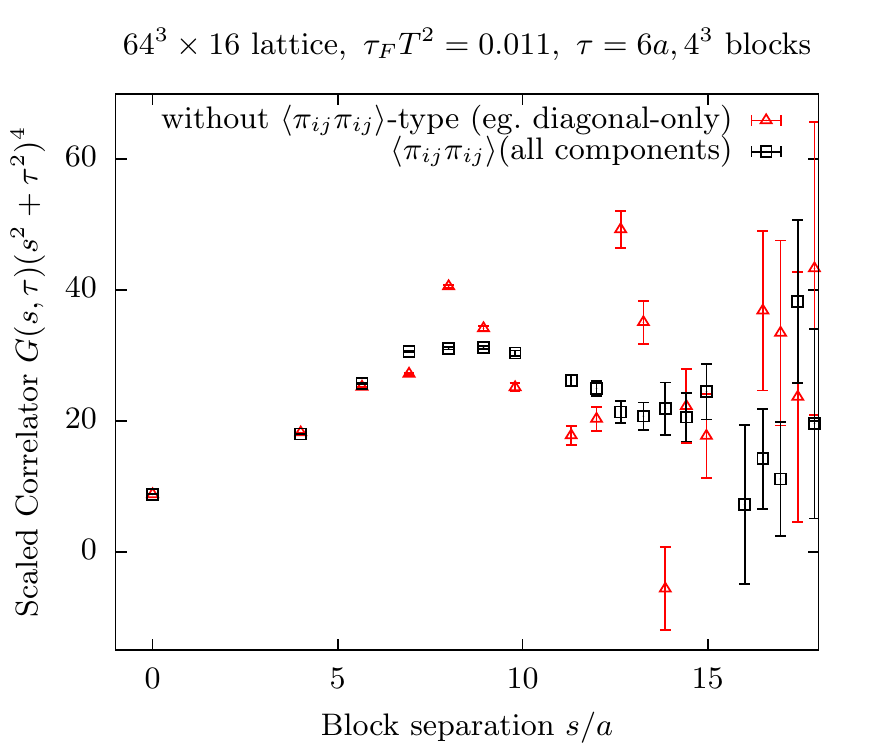}
\caption{Traceless spatial stress tensor (shear-channel) correlator between lattice blocks, at a fixed temporal separation, as a function of box separation, together with statistical error bars.
The black points contain all traceless stress tensor components, while the red data points contain only the diagonal-type contributions.
Some block separations only occur along lattice axes where the diagonal-type contributions are largest, while other block separations occur along lattice diagonals where some diagonal-type contributions are negative.
Hence, the red points jump around, while the black points follow a smooth curve until the statistical errors become large.
\label{fig:TTjump}}
\end{figure}

Each index combination of the $\langle \pi_{ij}(x,\tau) \pi_{ij}(y,0)\rangle$ correlator has a distinctive angular structure as a function of the direction of the $\vec x  - \vec y$ vector.
For instance, from reflection positivity we know that $\langle T_{xy}(\vec r) T_{xy}(0)\rangle < 0$ for $\vec r$ pointing along the $x$-axis or $y$-axis, but it is positive if $\vec r$ points along the $z$-axis or the line $x=y$.
In contrast, the $\langle (T_{xx}-T_{yy})(\vec r) (T_{xx}-T_{yy})(0)\rangle$ correlator is positive along each lattice axis but is negative along the $x=y$ line.
In our blocking procedure, certain block separations primarily sample blocks which are separated along lattice axes, while others sample the directions along lattice diagonals or other combinations.
Therefore, $T_{xy}T_{xy}$-type correlators will be larger for some blocks and smaller for others, while $T_{xx}-T_{yy}$-type correlators will show the opposite trend.
Including a single component or a subset of possible components leads to a correlation function which varies strongly with separation-direction and therefore jumps up and down as a function of the block separation.
This effect goes away if we include all traceless $ij$ combinations, which is therefore obligatory.  We illustrate this in Fig. \ref{fig:TTjump}, which shows the $\pi\pi$ correlation function as a function of block separation.

In general, the lattice renormalization constant $c_1$ is different for $T_{xx}-T_{yy}$ than for $T_{xy}$, because the rotational symmetry which relates them in the continuum is absent on the lattice \cite{Caracciolo:1989pt,Caracciolo:1991cp,Giusti:2015daa}.
We have only evaluated the renormalization constant for the former operator type.
However, the application of gradient flow should remove rotation-invariance violations in operator normalizations up to corrections suppressed by $\mathcal{O}(a^2/\tauf)$.
Therefore any effects from this operator normalization issue should be removed in our fixed-$\tauf$ continuum limit. 
A recent masters thesis%
\footnote{Jonas Winter, private communication}
explores both renormalization constants as a function of flow and finds that they are consistent with each other within 2\% error bars already for $\tauf / a^2 = 0.4$, the smallest value used here.

In order to remove the large-separation data and therefore its noise, it is necessary to fit the large-separation tail to a physically-motivated Ansatz.
The fitted value is then used instead of the data at those separations where the block-by-block signal-to-noise ratio is poor.
For our Ansatz we will use the leading-order perturbative behavior of the correlation function, accounting for time periodicity, gradient flow, and our blocking procedure.
In vacuum, the leading-order correlator of two field strength tensors is
\begin{widetext}
\begin{align}
    \label{vacFF}
    &\langle F^a_{\mu\nu}(r) F^b_{\alpha\beta}(0) \rangle
    = \frac{g^2 \delta_{ab}}{\pi^2 r^4}
    \left[ \delta_{\mu\alpha} \delta_{\nu\beta} - \delta_{\mu\beta} \delta_{\nu\alpha}
    \vphantom{\frac{2}{r^2}}
%    \right.    \nonumber \\ & \left.
     - \frac{2}{r^2} \left(
     r_\mu r_\alpha \delta_{\nu\beta} - r_\mu r_\beta \delta_{\nu\alpha} - r_\nu r_\alpha \delta_{\mu\beta}
    + r_\nu r_\beta \delta_{\mu\alpha}
     \right)
    \right].
\end{align}
Applying gradient flow to a depth $\tauf$ modifies this expression to
\cite{Eller:2018yje}:
\begin{align}
    \label{vacFFflow}
    &\langle G^a_{\mu\nu}(r) G^b_{\alpha\beta}(0) \rangle_{\tauf}
    = \frac{g^2 \delta_{ab}}{\pi^2 r^4}
    \left[ A(r,\tauf) \left( \delta_{\mu\alpha} \delta_{\nu\beta} - \delta_{\mu\beta} \delta_{\nu\alpha} \right)
    \vphantom{\frac{2}{r^2}}
%    \right.   \nonumber \\ & \left.
     + \frac{B(r,\tauf)}{r^2} \left(
     r_\mu r_\alpha \delta_{\nu\beta} {-} r_\mu r_\beta \delta_{\nu\alpha} {-} r_\nu r_\alpha \delta_{\mu\beta}
    {+} r_\nu r_\beta \delta_{\mu\alpha}
     \right)
    \right],
    \\ &
    A(r,\tauf) = 1 - \left(1+\frac{r^2}{8\tauf} \right) e^{-r^2/8\tauf}  \,,
%    \\ &
\qquad \quad
    B(r,\tauf) = -2 + \left[ 2 - 2 \frac{r^2}{8\tauf}
    + \left(\frac{r^2}{8\tauf} \right)^2\right]
    e^{-r^2/8\tauf} .
\end{align}
Note that this is a continuum, not lattice, expression; but when $\tauf /a^2 > 0.5$, the lattice-continuum difference for flowed correlators is small, and the use of a continuum limit at fixed flow depth based only on data which satisfies this criterion should avoid the need to include lattice spacing corrections as well.

Using these expressions, at finite $\tauf,\tau,|\vec r|$ and with periodic boundaries in the time direction, the leading-order stress tensor correlator summed over all transverse-traceless elements $\hat T_{ij} = T_{ij} - \delta_{ij} T_{kk}/3$ relevant for shear viscosity and for bulk viscosity are
%\begin{widetext}
\begin{align}
    \langle \hat T_{ij}(\vec r,\tau) \hat T_{ij}(0,0)\rangle_{\tauf}  \propto
    \sum_{n_1,n_2 \in \mathcal{Z}} &
    \frac{A(r_1) A(r_2)}{r_1^4 r_2^4}
    + \frac{A(r_1) B(r_2) + A(r_2) B(r_1)}{2 r_1^4 r_2^4}
    \nonumber \\ & 
    + \frac{B(r_1) B(r_2)}{6r_1^6 r_2^6}
    \left( 3 (r_1\cdot r_2)^2
    + \vec{r}^2 \left[ r_1^2+ r_2^2 - 4r_1\cdot r_2 + \frac{4}{5} \vec{r}^2 \right]\right),
    \\
    \langle T_{\mu\mu}(\vec r,\tau) T_{\nu\nu}(0,0) \rangle_{\tauf} \propto
    \sum_{n_1,n_2 \in \mathcal{Z}} &
    \frac{A(r_1) A(r_2)}{r_1^4 r_2^4} + \frac{A(r_1) B(r_2) + A(r_2) B(r_1)}{2r_1^4 r_2^4}
%    \nonumber \\ &
    + \frac{B(r_1) B(r_2)}{6r_1^6 r_2^6} \left(
    2 (r_1\cdot r_2)^2 + r_1^2 r_2^2 \right) \,,
\end{align}
where $r_1=(\tau+n_1 \beta,\vec r)$ and 
$r_2=(\tau+n_2 \beta,\vec r)$ are the 4-displacement with the temporal displacement shifted by independent integer multiples of the inverse temperature $\beta$.

\end{widetext}

\section{Temperature correction and tree level improvement}
\label{sec:improve}

\begin{figure*}[tbh]
\centerline{\includegraphics[width=0.5\textwidth]{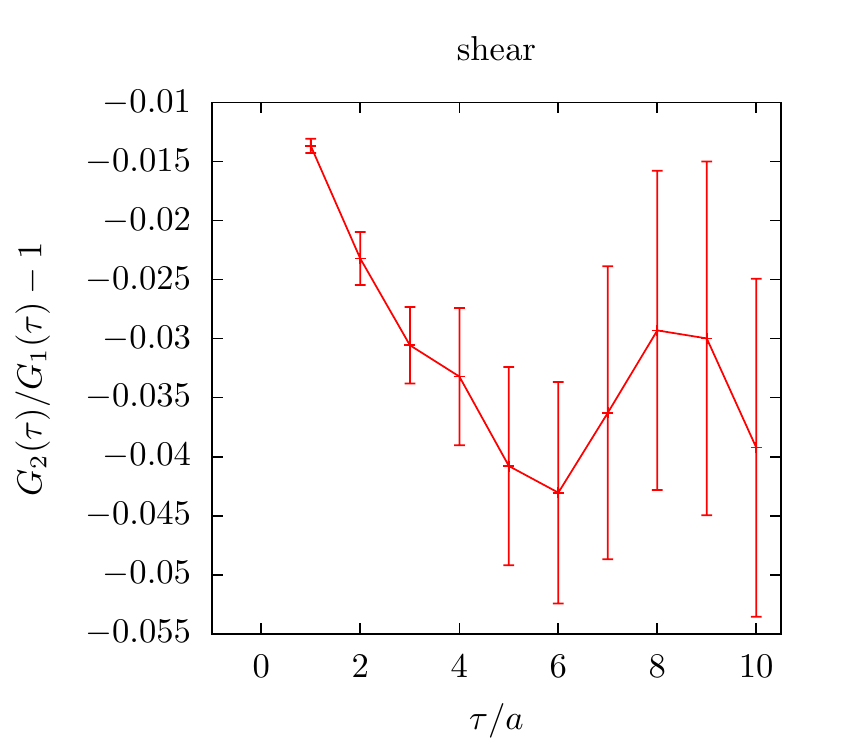}
\includegraphics[width=0.5\textwidth]{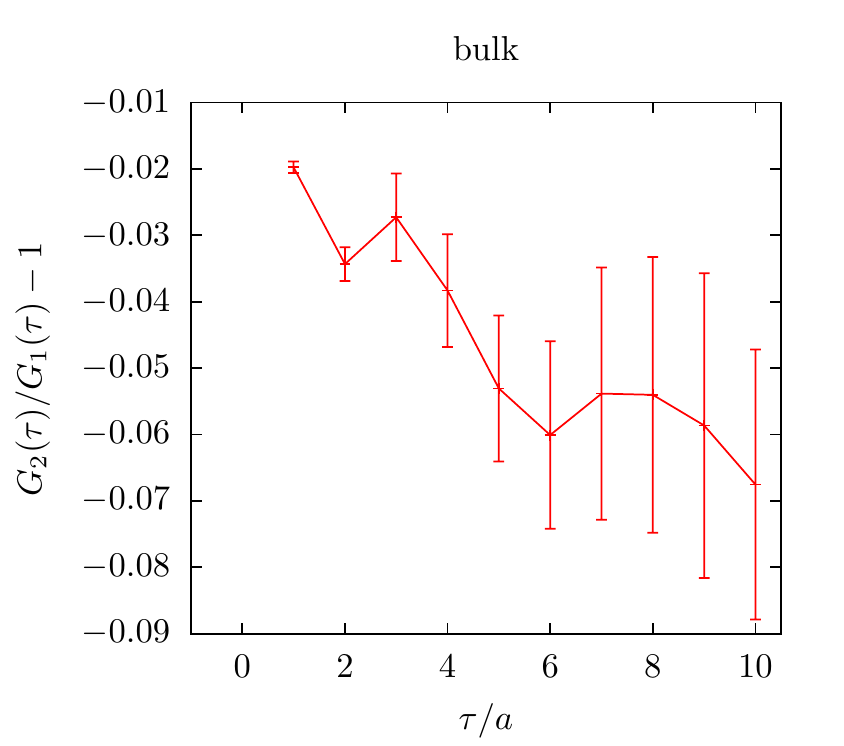}
}
\caption{The temperature correction when going from $\beta_1=7.035$ to $\beta_2=7.0767$ on an $80^3\times 20$ lattice, for shear (left) and bulk(right).
Each data point is found at the maximum flow time used in the flow-time extrapolation (see Sec.(\ref{sec:extrapolate})).
%The colored bands denote the error in the weighted average of all the data points.  % THERE IS NO COLORED BAND
}
\label{G1G2}
\end{figure*}

%\todoluis{should we maybe explain why we didn't choose the ``correct'' beta values in the first place?} \todomax{I think it is ok not to give more details.}

From Table~\ref{tab:lattice_setup} it can be seen that the temperatures are not exactly $1.5T_c$ on all lattices.
This setup is adopted for historical reasons \cite{Francis:2015daa, Ding:2021ise}, and the deviations of the temperature were only discovered after the correlators were measured.
The temperature differences, though small, must be accounted for when performing a continuum extrapolation.
Because the temperature differences are small and the lattices are fine enough that the continuum extrapolation is not very severe, we will content ourselves by evaluating the temperature dependence at the linearized level and at a single lattice spacing.
We then assume that the established temperature correction also applies at the other lattice spacings.
We choose to perform a linear temperature-dependence analysis on the lattice which has the largest deviation from $T=1.5T_c$, namely the $20\times 80^3$ lattice with $\beta\equiv \beta_1 = 7.035$ and $T = 1.4734 T_c$.
For this lattice, we choose a second $\beta$ value, $\beta_2 = 7.0767$, corresponding to $T = 1.5501 T_c$, and we repeat our correlation function studies on this lattice. 
%\textcolor{blue}{
Since the renormalized correlators contain two parts, namely the renormalization constants $c_1$ or $c_2$ and the bare correlators, the corrections for both parts should be considered.
The renormalization constants have been determined precisely in Sec.\ref{sec_renormalization} at $\beta$ values listed in Table~\ref{tab:lattice_setup}.
To obtain the one at $\beta=7.0767$ we linearly interpolate between $\beta=7.035$ and $\beta=7.192$.
We then calculate the renormalized correlators, denoted as $G_1$ and $G_2$ for the lower and higher temperature, respectively, by multiplying the bare correlations functions and the squared renormalization constants. We then evaluate the difference,
$-1+G_2(\tau) / G_1(\tau)$, representing the temperature dependence of the correlation function, as a function of $\tau$ and $\tauf$.
Statistical errors are computed using bootstrap sampling, and since $G_1,G_2$ arise from different ensembles, their errors are independent and can be propagated via Gaussian error propagation.

Figure~\ref{G1G2} shows the thermal correction for the largest gradient flow depth we use (and therefore the least noisy data).
The figure shows that the temperature effect is nearly $\tau$ independent except at the smallest $\tau$ values (which are contaminated by lattice effects).
Based on this result, we treat $-1+G_2/G_1$ as a function of $\tauf$ only, determining its value based on the weighted average of all the points at $\tau/a\geq 4$ at each $\tauf$.
As the figure shows, the thermal corrections are relatively small, considering that the temperature difference $1.5501-1.4734=0.0767 T_c$ is significantly larger than any of the individual deviations from $1.5 T_c$ shown in Table \ref{tab:lattice_setup}.
We will therefore use the determined slope $P=(G_2/G_1 - 1)/(T_2 - T_1)$, averaged over $\tau$ values, and apply it as a linearly interpolated correction to all data.
For instance, data at temperature $T$ can be interpolated to the temperature $T_0$ through $G(T_0) \simeq G(T) (1+ P (T_0-T))$.
A detailed analysis on the uncertainties in the temperature correction can be found in Appendix \ref{sec:error-T-correction}.
The appendix also presents an alternative model, which gives a consistent result.

Next, consider discretization effects associated with computing on a lattice rather than in continuous space.
To suppress the lattice discretization effects, we apply tree level improvement to the bare correlators.
Specifically, if we assume that the lattice correlation functions will deviate from the continuum ones in the same way as occurs at lowest-perturbative order, then we can remove this effect by rescaling by the ratio of leading-order continuum to lattice correlation functions
%This can be done by either improving the distance $\tau T$ of the correlators or directly improving the correlator itself. In this work we used the latter. We also crosschecked by applying the former and found the difference is negligible. 
%
%In the latter case one just multiplies the non-perturbative correlators measured on the lattice by a factor equal to the ratio of the leading-order continuum correlators to the leading-order lattice correlators (both without gradient flow)
\cite{Gimenez:2004me,Meyer:2009vj},
\begin{equation}
\label{ratio_tt}
G^{\mathrm{t.l.}}(\tau T)=G_{\rm lat}(\tau T)  
\,\cdot\, \frac{G^{\mathrm{LO}}_{\mathrm{cont}}(\tau T) }{G^{\mathrm{LO}}_{\mathrm{lat}}(\tau T) }.
\end{equation}
The leading-order continuum correlators in shear channel and  bulk channel can be found in~\cite{Meyer:2007ic,Meyer:2007dy}
\begin{equation}
\label{pert_cont}
\begin{split}
&\frac{G^{\mathrm{LO,shear}}_{\mathrm{cont}}(\tau T)}{T^5} = \frac{32d_A}{5\pi^2} \Big(f(x)-\frac{\pi^4}{72} \Big)\\
&\frac{G^{\mathrm{LO,bulk}}_{\mathrm{cont}}(\tau T)}{T^5} = \frac{484d_A}{16\pi^6}g^4 \Big(f(x)-\frac{\pi^4}{60} \Big),
\end{split}
\end{equation}
where $x=1-2\tau T$, $f(x) = \int_0^\infty ds~ s^4  \cosh^2(x s)/\sinh^2 s$ and $d_A=8$ counting the number of gluons. The leading-order lattice correlator for clover discretization is available in~\cite{Meyer:2009vj}. For better visibility we always normalize the tree-level improved correlators with a normalization correlator $G_{\rm{norm}}$ calculated at $\tauf=0$, where for shear channel we use $G_{\rm{norm}}\equiv G^{\mathrm{LO,shear}}_{\mathrm{cont}}$ and for bulk we use $G^{}_{\rm{norm}}\equiv G^{\mathrm{LO,bulk}}_{\mathrm{cont}}/g^4$.

\begin{figure*}[htb] 
\centerline{\includegraphics[width=0.5\textwidth]{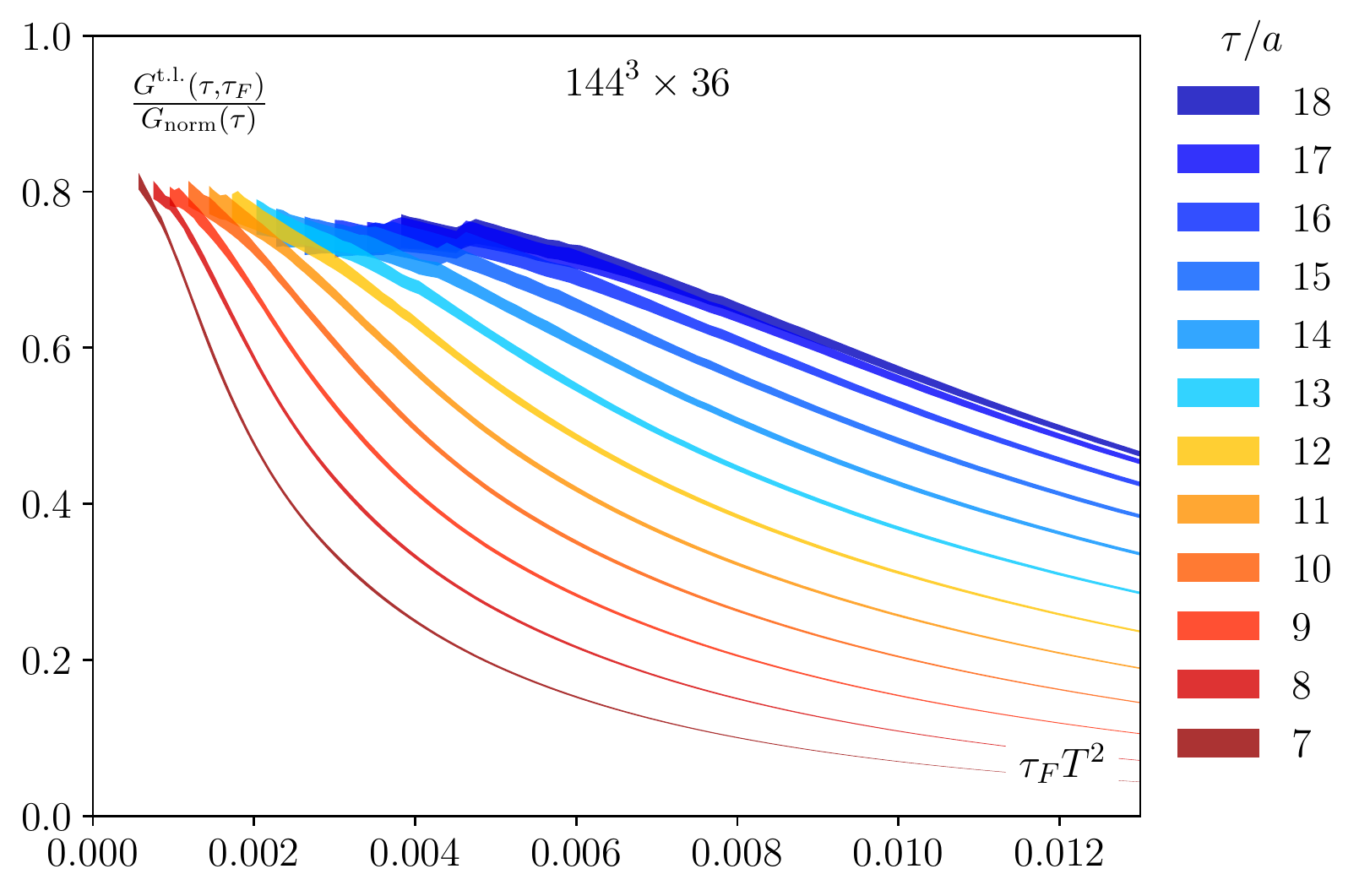}
\includegraphics[width=0.5\textwidth]{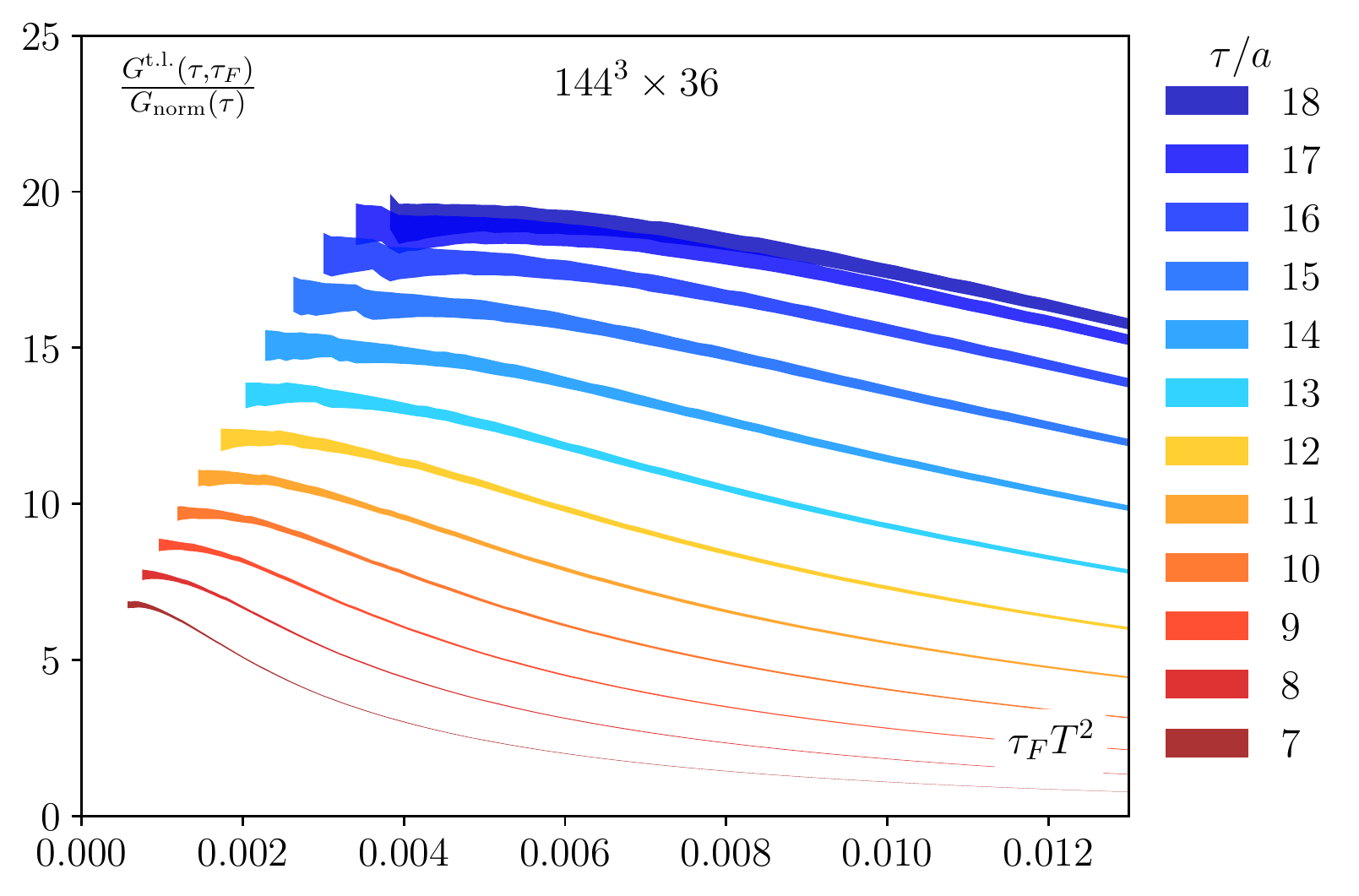}
}
\caption{Tree-level-improved EMT correlators in the shear channel (left) and bulk channel (right) normalized by the leading-order correlator on the $144^3\times 36$ lattice at different flow times.
(The tree-level correlator used for the normalization in the bulk channel is missing a factor of $g^4$, which explains the large ratio.)}
\label{Nt36corrs}
\end{figure*}

After temperature corrections, tree-level improvement and renormalization, in Fig.~\ref{Nt36corrs} we show the lattice correlators normalized by the free continuum correlators on $144^3\times 36$ lattice at different flow times, in both the shear and the bulk channels.
We have not plotted data down to small flow times because it has large errors.
We can see that as flow time increases the signal-to-noise ratio improves.
At very large flow times the signal is strongly modified by flow effects and we leave the regime where an extrapolation $\tauf \to 0$ can be performed.

\section{Double extrapolation}
\label{sec:extrapolate}

\begin{figure*}[htb] 
\centerline{\includegraphics[width=0.5\textwidth]{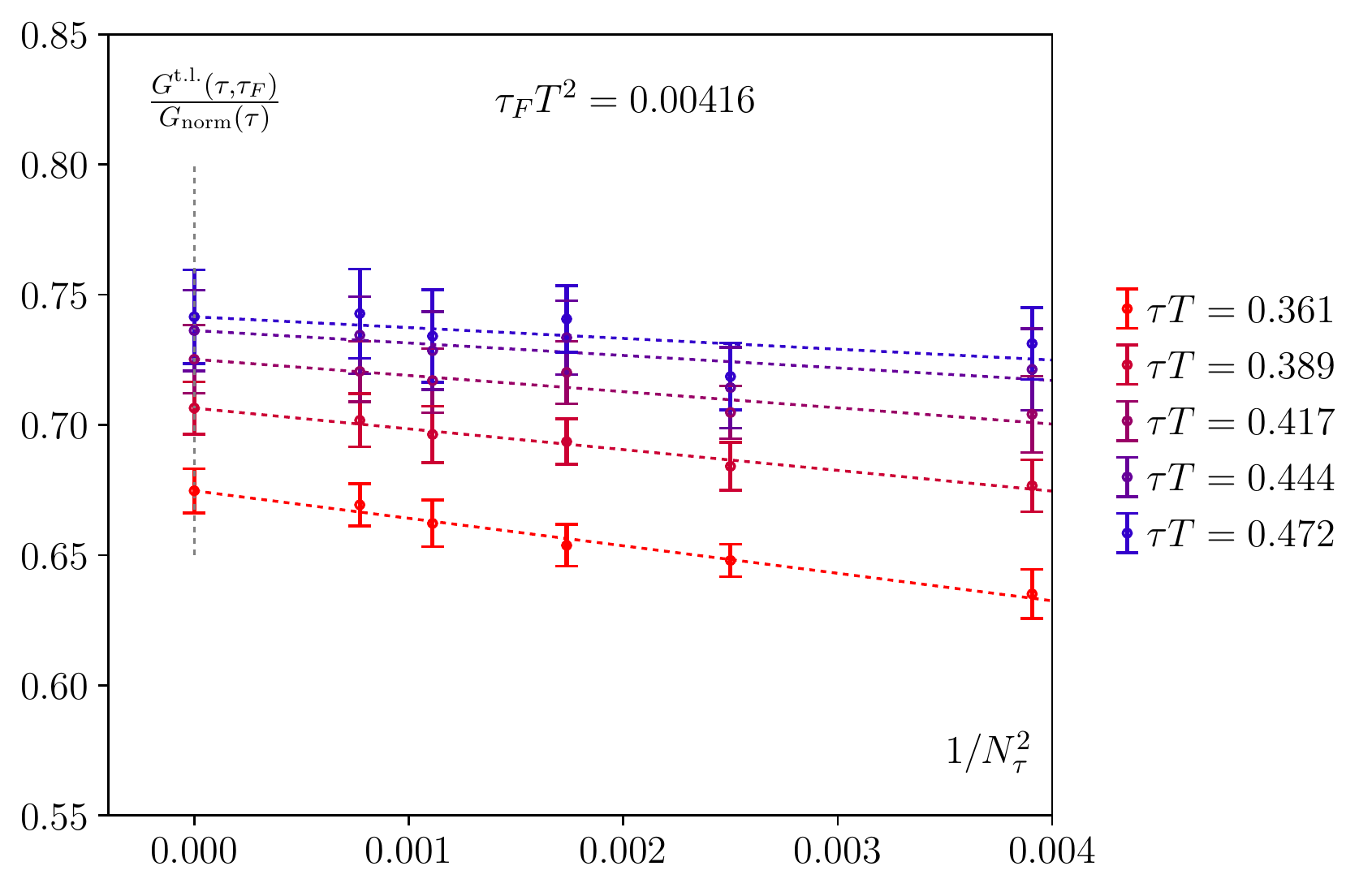}
\includegraphics[width=0.5\textwidth]{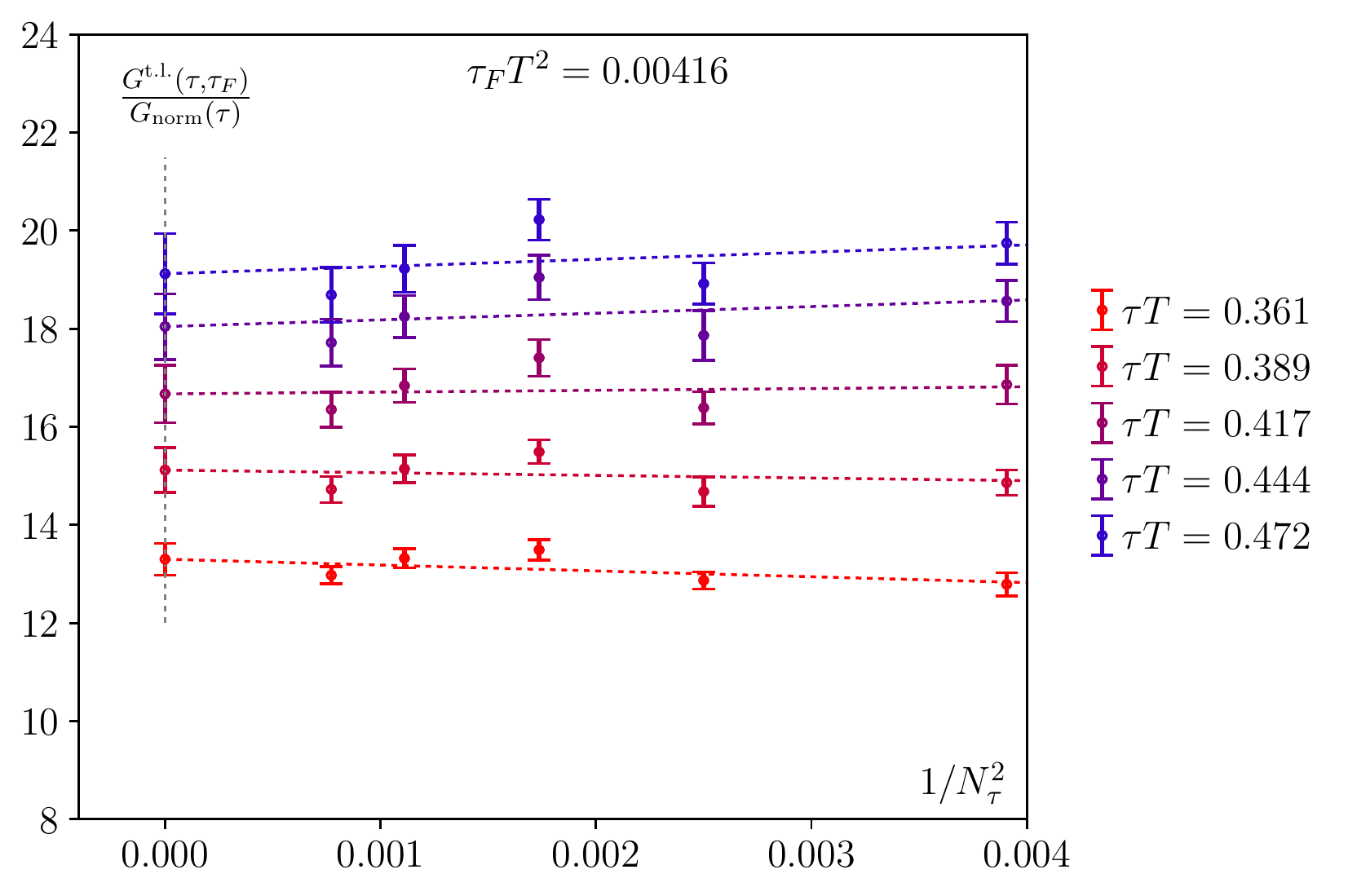}
}
\caption{The continuum extrapolation of EMT correlators in shear channel (\textit{left}) and bulk channel (\textit{right}) at flow time $\tauf T^2=0.00416$, fit using Eq.~(\ref{cont_Ansatz}).
The error bars on data points are statistical; the errors on the extrapolated values are the uncertainties from the extrapolated fit.
}
\label{cont-extrap}
\end{figure*}

The double extrapolation contains two steps: first we perform the continuum extrapolation $a \to 0$, and then we perform a flow-time-to-zero extrapolation.
As we pointed out in Ref.~\cite{Altenkort:2020fgs}, this has the advantage that the continuum extrapolation eliminates terms of form $a^2 / \tauf$, so that the $\tauf$ extrapolation will consist only of positive powers.
Before the continuum extrapolation, the correlators on coarse lattices have to be interpolated to the separations of the finest lattice, for details see, for example, references \cite{Altenkort:2020fgs, Altenkort:2020axj}.
In the continuum extrapolation we use the Ansatz 
\begin{align}
\label{cont_Ansatz}
    \frac{G^\textrm{t.l.}(N_\tau) }{G_{\rm{norm}}(N_\tau)}
     =  m \cdot N_\tau^{-2} + b,
\end{align}
because the lattice action has leading discretization errors of order $a^2$. Here $m$ and $b$ are fit parameters that can be different for each temporal separation and flow time.
The continuum estimates for the (normalized) correlators are given by $b \equiv G_\mathrm{cont}/G_{\rm{norm}}$. 

%\todoguy{We don't mention at all that we have to interpolate values on coarser lattices.  At least this should be mentioned and we should refer to a paper where the details appear!}

\begin{figure*}[htb] 
\centerline{\includegraphics[width=0.5\textwidth]{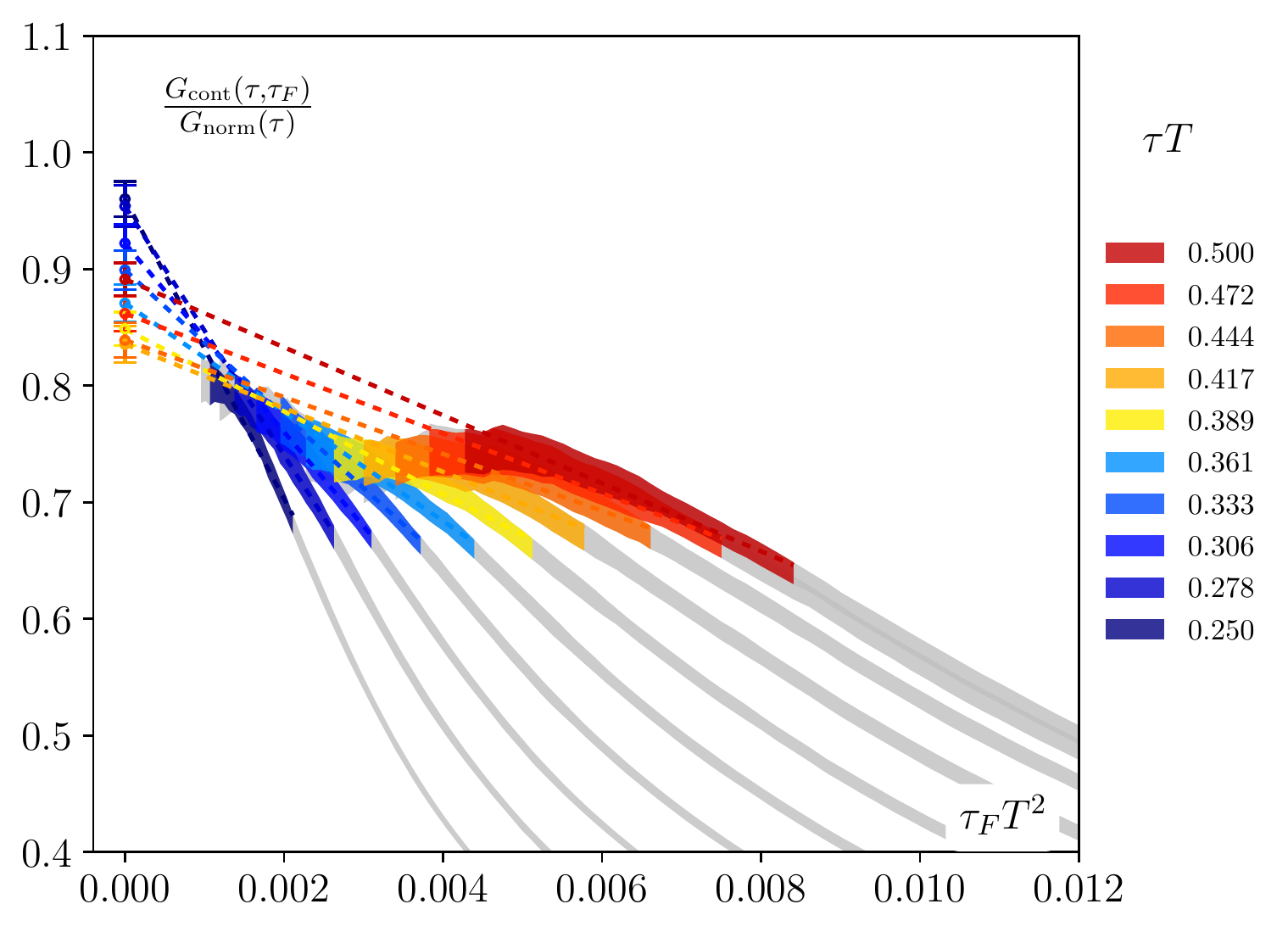}
\includegraphics[width=0.5\textwidth]{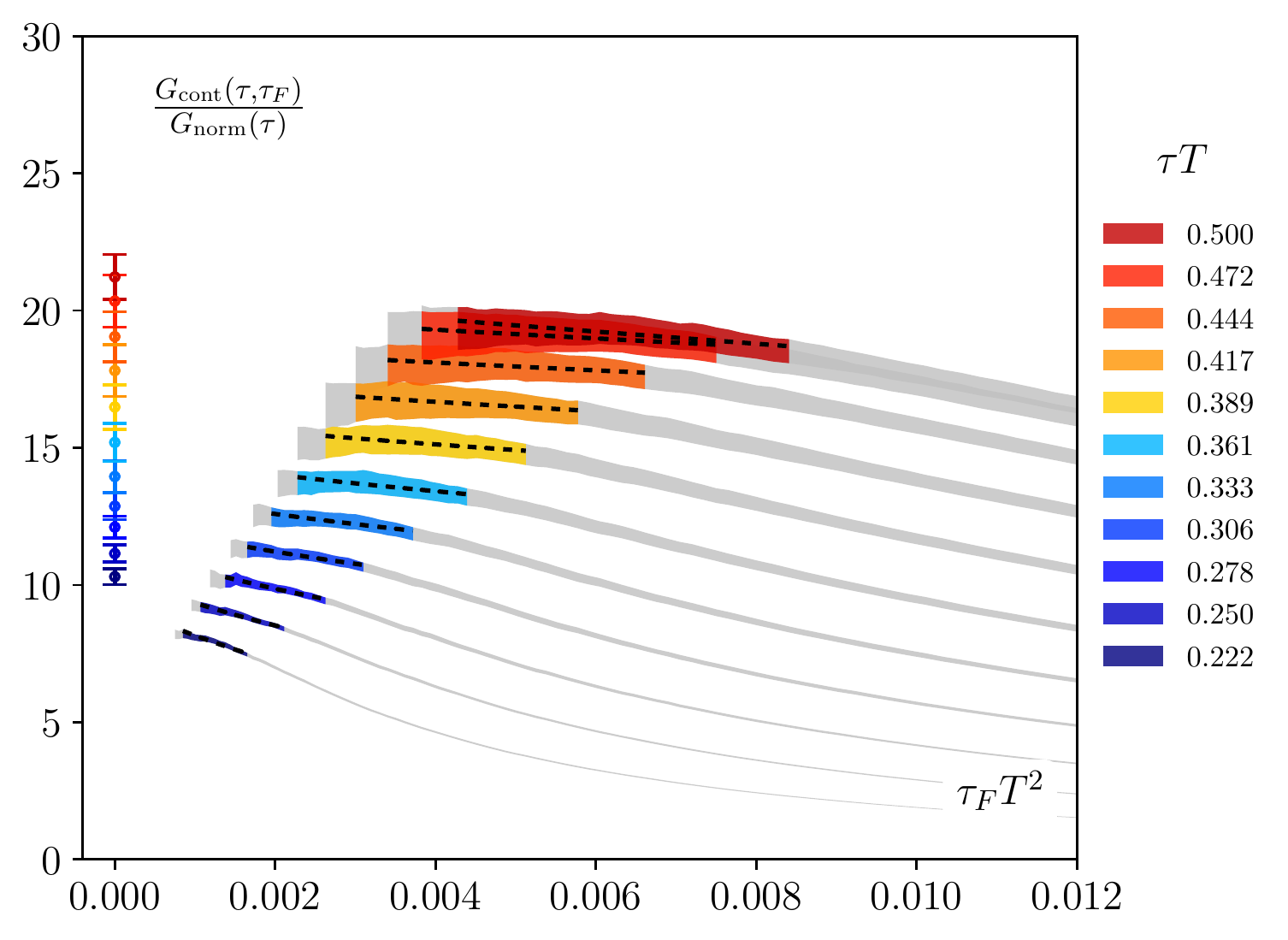}
}
\caption{The $\tauf\rightarrow 0$ extrapolation of continuum-extrapolated EMT correlators in the shear channel (\textit{left}) and bulk channel (\textit{right}). 
}
\label{tauF-extrap}
\end{figure*}

Figure~\ref{cont-extrap} shows how good the fit Ansatz, Eq.~(\ref{cont_Ansatz}), works at an intermediate flow time $\tauf T^2=0.00416$.
We can see for the bulk channel that in some cases the fit is poor in the sense that $\chi^2/\mathrm{dof} > 1$.
Our procedure is to enlarge the error bars by $\sqrt{\chi^2/\mathrm{dof}}$ in these cases. 
After the continuum extrapolation we collect the continuum estimates for each flow time and show them in grey bands in Fig.~\ref{tauF-extrap}.

\begin{figure*}[tbh]
%\centerline{\includegraphics[width=0.5\textwidth]{./figures/shear_corr.pdf}
%\includegraphics[width=0.5\textwidth]{./figures/bulk_corr.pdf}
%}
\centerline{
\includegraphics[width=0.5\textwidth]{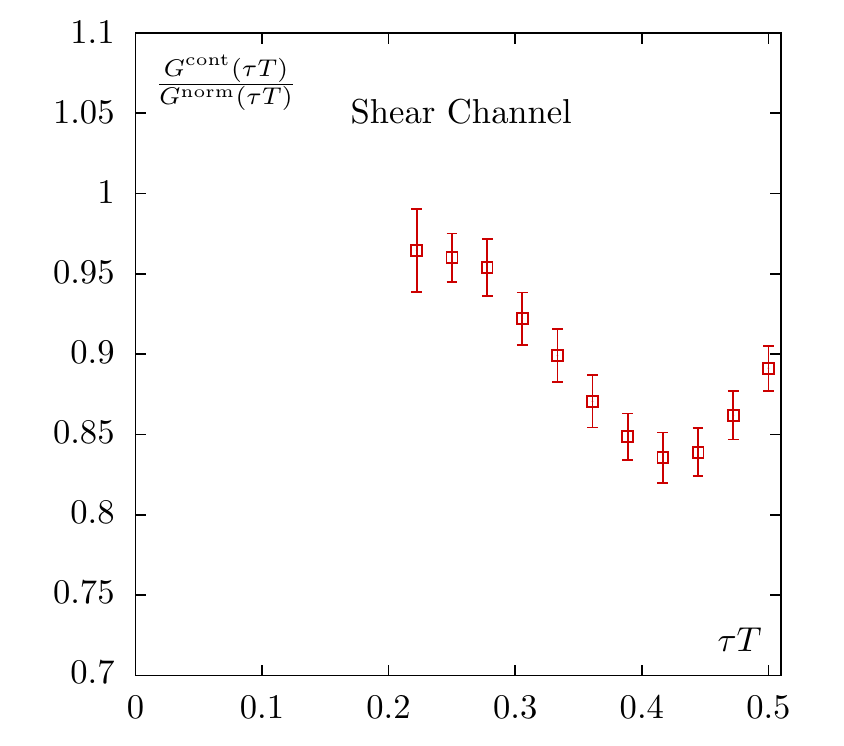}
\includegraphics[width=0.5\textwidth]{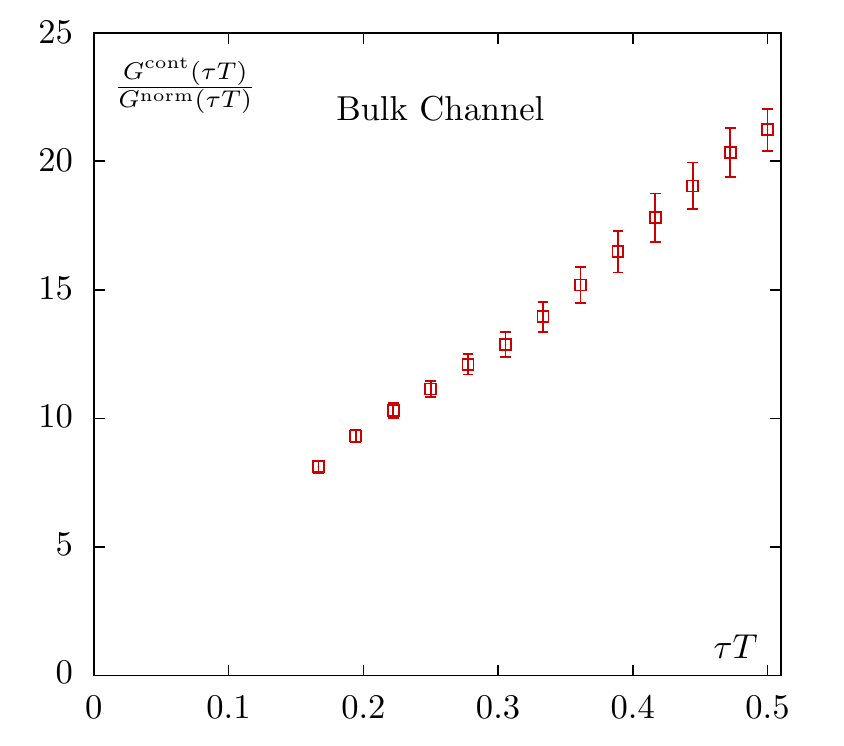}
}
\caption{Double-extrapolated correlators in the shear channel (\textit{left}) and bulk channel (\textit{right}).
Note that $G^{\mathrm{norm}}(\tau T)$ in the bulk channel is missing a factor of $g^4$, which explains the size and possibly the slope of the resulting correlator ratio.
}
\label{final_corrs}
\end{figure*}

Now we consider the $\tauf\rightarrow 0$ extrapolation.
To perform the extrapolation, we need to understand the functional dependence on $\tauf$, and we need to determine over what range of $\tauf$ values to perform the extrapolation.
For general values of $\tauf / \tau^2$, the correlator is a complicated function of this ratio, in some cases even taking on a different sign than the small-$\tauf$ value \cite{Eller:2018yje}.
However, if $\tauf / \tau^2$ is small, then as discussed near the end of Section \ref{sec:gradflow}, we expect the flowed stress tensor to be described in terms of an operator product expansion, with the leading coefficient equaling the stress tensor and with higher-dimension operators suppressed by powers of $\tauf$.
As a result, in this regime the small-$\tauf$ expansion of the correlation function should approach $\tauf \to 0$ with polynomial-in-$\tauf$ corrections.
(We will ignore possible anomalous dimensions in this discussion.)
%\todomax{Can we simply ignore possible anomalous dimensions here? If I remember correctly, Guy and I looked at them for EE-correlator and I think we concluded that they are not always small effects.}
%\todoguy{Hi Max.  I think you are thinking of something different -- if you ignore anomalous dimensions in the gauge-field correlator, then the linear term discussed below is zero.  The question here is whether it is linear or slightly-tilted-from-linear.  If we want to allow it to be a power different from linear, that's another fitting parameter . . . }

The more fitting coefficients we use, the larger the errors in the resulting fit.
Therefore we want to avoid using two extrapolation coefficients,
e.g., a fit of form $G(\tauf/\tau^2) = A + B \tauf/\tau^2 + C \tauf^2/\tau^4$.
And if we use a wide enough data range that the $\tauf^2/\tau^4$ coefficient is really relevant, then there is a danger that we also need still higher-order coefficients.
Therefore, we will restrict ourselves to a region where the total variation in $G(\tauf/\tau^2)$ appears to be at most 20\% from its extrapolated value.
In this range, within the few \% accuracy which is our goal, we expect that a linear extrapolation, e.g., $G(\tauf/\tau^2) = A + B \tauf/\tau^2$, should be sufficient.
Based on our previous experience with the topological density operator
\cite{Altenkort:2020axj}, we expect that a fitting range out to $\sqrt{8\tauf^{\mathrm{max}}} = 0.5220\tau$ should remain in this small-correction regime.
We will fit a range of $\tauf$ from this maximum down to half this value, because the correlator becomes so noisy at smaller $\tauf$ that extending the range further is not helpful.
In addition, to prevent lattice spacing effects of form $a^2/\tauf$, we restrict to values with $\tauf/a^2 \geq 0.4$ as already discussed.
For small $\tau$ values this constraint excludes too much of the $\tauf$ range over which we want to extrapolate, which prevents us from determining the correlator at small temporal separations.
The resulting correlators within the range $[0.5\tauf^{\mathrm{max}} , \tauf^{\mathrm{max}}]$ are shown as colored bands in Fig.~\ref{tauF-extrap}. 

For the extrapolation of the bulk viscosity correlators we have taken a slightly different approach, based on the work of 
\cite{Suzuki:2013gza, Makino:2014taa,Suzuki:2021tlr}.
A recent three-loop calculation of the flow-dependence of the EMT trace suggests a finite-$\tauf$ fitting function of form \cite{Suzuki:2021tlr}
\begin{equation}
\theta(\tauf)=\Big{(}1-c \big{(}\frac{g^2(\mu(\tauf))}{(4 \pi)}\big{)}^{3}\Big{)} \theta(\tauf=0), 
\label{resquenchS} 
\end{equation}
where $c$ and $\theta(\tauf=0)$ are fit parameters.
Since what we measured in this study is the correlators of $\theta$, we take the square root of the correlators and fit it to Eq.~(\ref{resquenchS}). The fitted curves are shown as dashed black lines in Fig.~\ref{tauF-extrap} and the extrapolated correlators are shown as colored points at $\tauf T^2=0$. It can be seen that the fit function is almost linear, indicating that a fit to an Ansatz linear in flow time (as used in \cite{Altenkort:2020fgs,Altenkort:2020axj}) would give similar results.
%\textcolor{blue}{
Appendix \ref{sec:error-double-extrap} presents more details on both the continuum and the small flow-time extrapolations. The double extrapolated correlators in both channels are shown in Fig.~\ref{final_corrs}.
%}

%As for the shear channel, due to the divergence of the renormalization constant at zero flow time, \wave{an extrapolation using a power function of the running coupling is tricky to handle}. So we choose to use a linear function in flow time, which should be valid at the lowest order, \wave{and is confirmed by our data}.The final double-extrapolated correlators are shown in Fig.~\ref{final_corrs}.
%\todoguy{I want to make sure that I understand this right.
%We consider a fit as suggested by Suzuki, we find it's almost the same as a linear fit, we feel that there are some complications with doing the fit the Suzuki way, and so we do a linear fit after all.
%Is that right?
%If it is, I propose that we just simply say that we did a linear fit, and skip the whole discussion here.}

\section{Spectral analysis}
\label{sec:spectrum}

This section is devoted to the spectral extraction from the extrapolated correlators. We first reconstruct the spectral function using $\chi^2$-fits with models based on perturbative calculations and then determine the viscosities using the Backus-Gilbert (BG) method \cite{Backus1968}.

The spectral reconstruction performed here is mathematically ill-posed \cite{hadamard1923lectures}. One feature of this is the difficulty in quoting a robust spectral function since uniqueness of any solution is $a$ $priori$ not given. 

In the case of the spectral analysis via fit this issue presents itself as the difficulty in finding a global, well-determined minimum. 
In principle, if the ``correct'' Ansatz were known, with enough data points and without considering any noise the analysis should yield a global minimum in the $\chi^2$-plane. Without this knowledge and with noise included, however, this minimum is less well determinable and a fit often yields $\chi^2$-values that are not very sensitive to the parameter choices. Consequently it becomes difficult to choose with confidence which solution and Ansatz is the best description. In the following we address this difficulty by augmenting our study with a spectral analysis using a method that does not rely on an Ansatz $per$ $se$ in form of the BG method.

\subsection{Spectral function from model fits}

According to Eqs.~(\ref{kubo1}) and (\ref{kubo2}), the viscosities are proportional to the slope of the spectral function at zero frequency. But the large frequency part also contributes considerably to the correlators and they
can be computed perturbatively. For the shear channel the large frequency part has been computed both at leading order (LO) and at next-to-leading order (NLO) \cite{Zhu:2012be},
\begin{align}
\begin{split}
\label{rhoLO}
\rho_{\rm{shear}}^{\mathrm{LO}}(\omega)= & 
 \frac{d_A \  \omega^4}{10\pi}
 \coth\Big{(}\frac{\omega}{4T}\Big{)},\\
\rho_{\rm{shear}}^{\mathrm{NLO}}(\omega)= & \rho_{\rm{shear}}^{\mathrm{LO}} (\omega) -
 4 d_A \omega^4
 \coth\Big{(}\frac{\omega}{4T}\Big{)} \frac{g^2(\bar{\mu})N_c}{(4\pi)^3} \\
 &\times \biggl[
    \frac{2}{9} 
     + \phi^{\eta}_T(\omega) 
    \biggr] .
 \end{split}
\end{align}
Note that our definition of the spectral function differs from
that in Ref.~\cite{Zhu:2012be} by a relative minus sign.
Here $d_A = N_c^2-1=8$ is the dimension of the adjoint representation.
In the region of $\omega\ll \pi T$, the one-loop running coupling can be fixed via the `EQCD' renormalization point \cite{Kajantie:1997tt}
\begin{align}
\label{mu_opt_from_T}
 \ln\left( \bar{\mu}^{\mathrm{opt}(T)} \right) \equiv \ln\left( 4\pi T\right) -\gamma_{\mathrm{E}} -\frac{1}{22} \,.
\end{align}

\begin{figure*}[tbh]
\centerline{\includegraphics[width=0.5\textwidth]{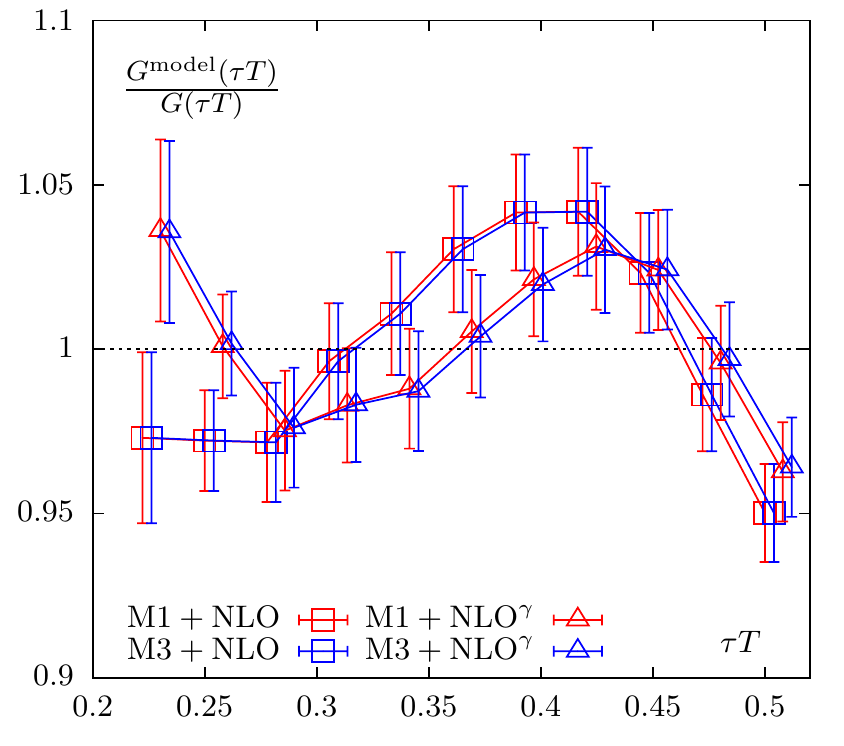}
\includegraphics[width=0.5\textwidth]{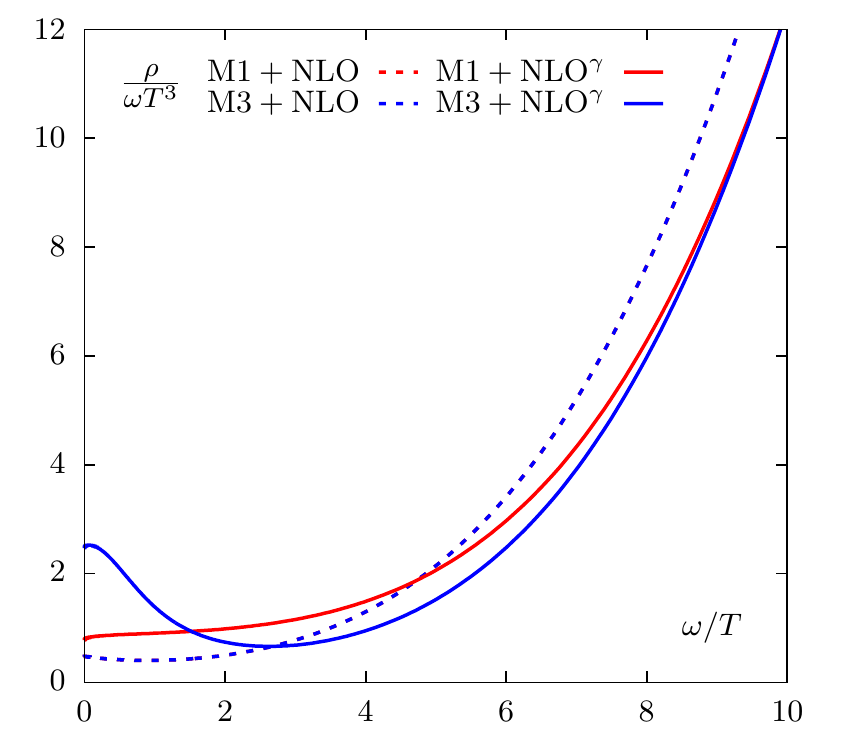}
}
\caption{The comparison of fit correlators and lattice correlators (left) and the fit spectral function in the shear channel. In M3 the width of the Lorentzian peak $C$ has been fixed to 1.
}
\label{fit_shear_corr_spf}
\end{figure*}

\begin{figure*}[tbh]
\centerline{\includegraphics[width=0.5\textwidth]{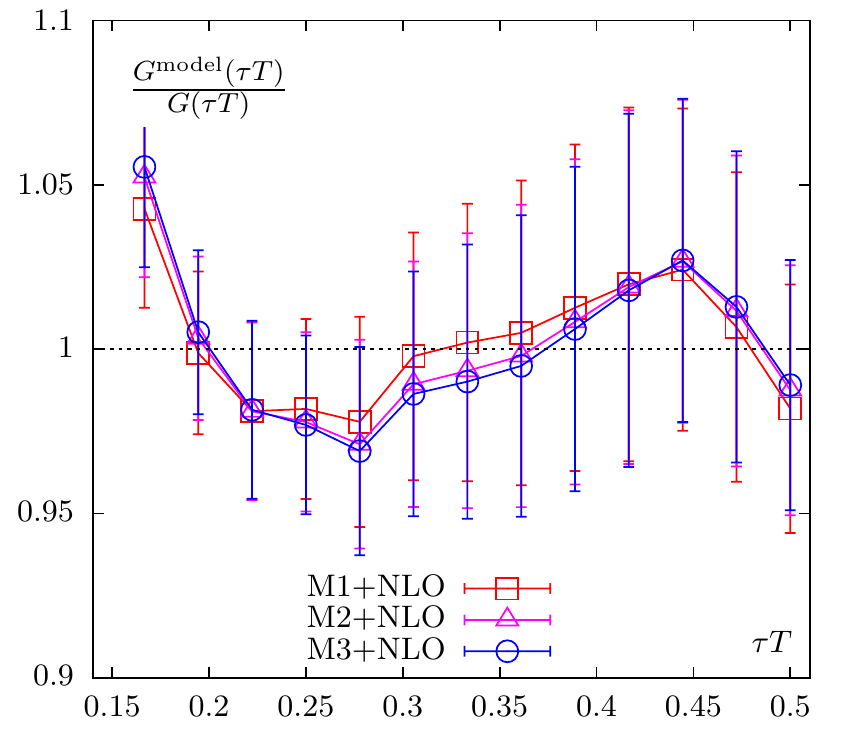}
\includegraphics[width=0.5\textwidth]{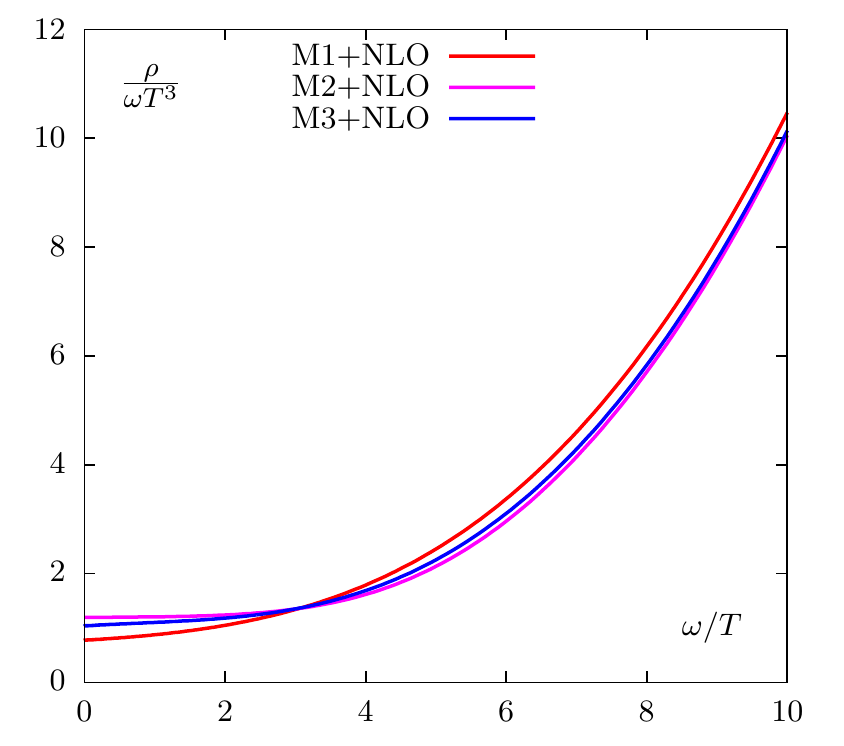}
}
\caption{The comparison of fit correlators and lattice correlators (left) and the fit spectral function in the bulk channel.
}
\label{fit_bulk_corr_spf}
\end{figure*}

Using this relation the coupling is fixed to the value $g^2\left( \bar{\mu}^{\mathrm{opt}(T)}\right)=2.2346$ at $T=1.5T_c$, where we use an updated relation $T_c=1.24\Lambda_{\overline{\mathrm{MS}}}$~\cite{Francis:2015lha}.
For large $\omega$, due to the lack of explicit logarithms of the renormalization scale in Eq.~(\ref{rhoLO}), a natural choice is given by $\bar{\mu}^{\mathrm{opt}(\omega)}=\omega$ \cite{Zhu:2012be}. Combining the above two conditions a switching point for the renormalization scale at  $\omega/T=2.146 \pi$ can be found. The  dimensionless function $\phi_T^{\eta}(\omega)$ was first determined in Ref. \cite{Zhu:2012be} but with a computational error, which was found in reference \cite{Vuorinen:2015wla}. In \cite{Vuorinen:2015wla} another term from HTL resummation was introduced. Such a term only affects small frequencies and we do not include it in our spectral analysis, as we do not expect HTL to be reliable at the nonperturbative regime of small frequencies.

For the bulk channel the LO
and NLO spectral function are also available \cite{Laine:2011xm} 
\begin{equation}
\begin{split}
\label{rho_pert}
\rho_{\rm{bulk}}^{\mathrm{LO}}(\omega)= & 
 \frac{d_A c^2_{\theta}\  \omega^4g^4}{4\pi}
 \coth\Big{(}\frac{\omega}{4T}\Big{)},\\
\rho_{\rm{bulk}}^{\mathrm{NLO}}(\omega)= & \rho_{\rm{bulk}}^{\mathrm{LO}} (\omega) +
 d_A c^2_{\theta}\ \omega^4
 \coth\Big{(}\frac{\omega}{4T}\Big{)} \frac{g^6(\bar{\mu})N_c}{(4\pi)^3} \\
   &\times \biggl[
       \frac{22}{3} \ln\frac{\bar{\mu}^2}{\omega^2} + \frac{73}{3} 
     + 8\, \phi^{ }_T(\omega) 
    \biggr],
 \end{split}
\end{equation}
where $c_{\theta}\approx -b_0/2-b_1 g^2/4$, $b_0=\frac{11N_c}{3(4\pi)^2}$ and $b_1=\frac{34N_c^2}{3(4\pi)^4}$. $\phi_T(\omega)$ can be found in \cite{Laine:2011xm}. At LO, the running coupling can not be fixed. For simplicity we fix it to the one at the switching point at NLO. One can also fix it to another point, however this will not have effect 
%\todoluis{the rescaling factor is the same for all $\omega$, right? that is not the same as running coupling}
%\todohaitao{what I wrote is misleading. I meant fixing to one point doesn't make difference than fixing it to another point, as they only give an overall factor.}
on our spectral reconstruction as we shall see later there will be a rescaling factor to account for this uncertainty. At NLO, for $\omega \gg \pi T$ the optimization of the scale $\bar{\mu}$ and the running coupling can be determined\cite{Laine:2011xm}
\begin{align}
\ln\left( \bar{\mu}^{\mathrm{opt}(\omega)} \right) \equiv \ln\left(  \omega \right)  -\frac{73}{44} \,.
\label{mu_opt_from_omega}
\end{align}
In the opposite regime one should use Eq.~(\ref{mu_opt_from_T}). Equating Eq.~(\ref{mu_opt_from_T}) to Eq.~(\ref{mu_opt_from_omega}) leads to a switching point $\omega/T=11.276 \pi$. For an arbitrary $\omega$ the larger optimization scale from the two equations should be used.

The infrared behavior of the spectral function is not known $a$ $priori$, and must be modeled.
In previous work \cite{Altenkort:2020axj} we have considered several proposed IR behaviors, generally finding that the data is not very restrictive between different IR Ans\"atz choices.
In this work we will consider one model with an infrared ``peak'' and perturbative UV behavior, and two ``peak-free'' models in which the IR behavior is linear in $\omega$, the UV behavior is perturbative, and the spectral function increases continuously between them,
%a single, rather simple model for the IR behavior, which is that the spectral function is linear in frequency in the IR, and we will connect this behavior to the LO and NLO UV behavior in one of three ways:
%Our strategy is to join the first-order hydrodynamic spectral function at small frequencies ( linear in frequency)  with the perturbative part, denoted as $\rho_{\mathrm{pert}}$, which can be the 
%either the LO or the NLO results \cite{Shifman:1978bx}. Since the fixing of the scale introduces systematic uncertainty whose size is hard to estimate, an overall scaling parameter $B$ is introduced to account for the uncertainties in the renormalization of the perturbative part. To connect these two parts we consider three simple ways of interpolation 
%\todohaitao{uhh we need discussions I think}
\begin{align}
\label{models}
\mathrm{M1}: \frac{\rho(\omega)}{\omega T^3}=
& \frac{A}{T^3}+ B\frac{\rho_{\mathrm{pert}}(\omega)}{\omega T^3},\\ \nonumber
\mathrm{M2}: \frac{\rho(\omega)}{\omega T^3}=
&\sqrt{\left(\frac{A}{T^3}\right)^2
+\left(B\frac{\rho_{\mathrm{pert}}(\omega)}{\omega T^3}\right)^2},\\ \nonumber
\mathrm{M3}: \frac{\rho(\omega)}{\omega T^3}=
&\frac{A}{T^3}\frac{C^2}{C^2+(\omega/T)^2}+B\frac{\rho_{\mathrm{pert}}(\omega)}{\omega T^3}
\end{align}
Here $B$ is a coefficient allowing for a rescaling of the perturbative result, and $A$ is the size of the IR contribution, which determines the transport coefficient of interest.
In the first model, we consider a simple sum of an IR and a UV behavior; in the second, we consider a smooth switch-over between IR and UV behavior.
In the third model, the IR behavior is a Lorentzian with width parameter $C$.
For simplicity, we have fixed the width parameter $C$ to unity, but we also explored other values and we find a rather weak dependence of the fit quality on the choice.
We will use the range of fit values for $A$ between these models as an estimate of the value and uncertainty in the viscosity, though realistically the true spectral function may look different than any of our models and this introduces a potentially large systematic uncertainty in our final result.
In addition, for the bulk-viscous channel, there is a known constant contribution arising from the dependence of $T_{\mu\mu}$ on the energy density and on the fluctuations in the system energy.
Specifically, the spectral function is known to possess a delta function at zero frequency, equal to
$\rho/\omega T^3 = \pi\frac{E+P}{T^3}\frac{(3c_s^2-1)^2}{c_s^2}\delta(\frac{\omega}{T})$.
Equivalently one can subtract an $\tau$-independent constant of corresponding size from the Euclidean correlation function.
We adopt the values $\frac{E+P}{T^3}=5.098$ and $c_s^2=0.2848$ that can be calculated from \cite{Giusti:2016iqr}.

For the bulk channel our fit has two parameters on 13 data points, leaving 11 degrees of freedom.
The leading-order fit shows a poor $\chi^2$/dof, with values of 3.9, 5.4 and 6.3 for $\mathrm{M1}$,  $\mathrm{M2}$, and $\mathrm{M3}$ with $C=1$.
But using the NLO spectral function returns a good fit, with $\chi^2$/dof of 0.4, 0.5 and 0.6.
This suggests that the NLO corrections and in particular the running of the coupling improve the estimation significantly and brings it close to our non-perturbative determination.
The resultant $\zeta/T^3$ is $0.086(0.008)$, $0.133(0.010)$, and $0.303(31)$  for M1, M2, and M3($C=1$), respectively.

For the shear channel we find that when using the LO spectral function the $\chi^2$/dof is 4.1, 3.99 and 3.98, and for the NLO spectral function it is 3.7, 4.8 and 3.66, respectively.
This indicates that both LO and NLO calculations fail to capture our nonperturbative results for the Euclidean correlator.
This indicates that the true form of the spectral function is something more complicated than our relatively simple proposals in Eq.~(\ref{models}).

%calculation, which might have more/different structures in the spectral function, consistent with the bad convergence of shear viscosity in perturbative calculations observed in \cite{Ghiglieri:2018dib}. 

As one attempts to capture possibly missing structure, we have considered amending the UV part of the spectral function with an anomalous dimension, namely changing Eq.~(\ref{rhoLO}) by replacing
%To account for the potential change that might be needed for a correct spectral function, we further introduce a anomalous dimension to the perturbative spectral functions, namely in Eq.~(\ref{rhoLO}) we replace 
$\omega^4$ with $\omega^{4+\gamma}$.
With this modification we find that the $\chi^2$/dof becomes $\sim$2.0-2.1 for all models, both for the LO and the NLO spectral function.
The returned value of the viscosity, with statistical errors, is $\eta/T^3=0.84(0.14)$, $1.10(0.14)$ for LO and $0.77(0.16)$ and $1.09(0.15)$ for NLO, all using the first two models.  Model M3 with $C=1$ using NLO and an anomalous dimension returns $\eta/T^3 = 2.46(54)$.
%\textcolor{blue}{something about M3 for shear, again, lower bound $\sim$1.07(0.19) from $C=5$,}
Using an anomalous dimension improves the fit, but $\chi^2$/dof of 2 with eight degrees of freedom still represents a rather poor fit. We show the ratio of fit correlators to the lattice data, and the resulting spectral functions in Fig.~\ref{fit_shear_corr_spf} and Fig.~\ref{fit_bulk_corr_spf} for the shear and bulk channel, respectively. 
It would be interesting to explore other models for the IR behavior and to see if any such model can improve the quality of our fit.

%\wave{Since $\chi^2$/dof=2 is still not perfect}, which we are not able to decrease more without further knowledge about the structure of the spectral function, we would like to crosscheck the reliability of our fits by applying Backus-Gilbert method given in the next section.

\subsection{Spectral function from Backus-Gilbert method}

\begin{figure*}[tbh]
\centerline{\includegraphics[width=0.5\textwidth]{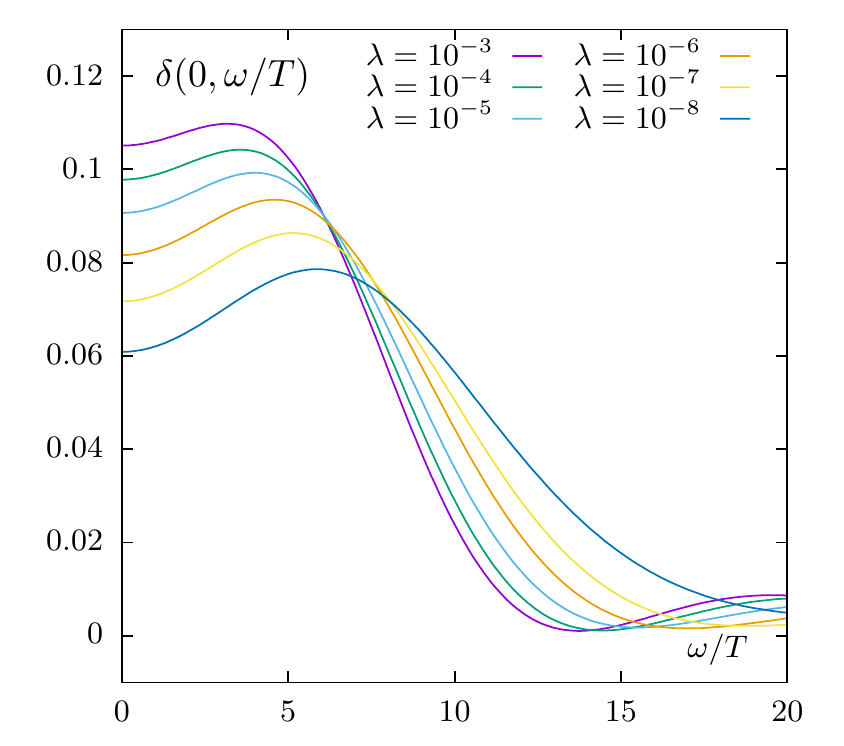}
\includegraphics[width=0.5\textwidth]{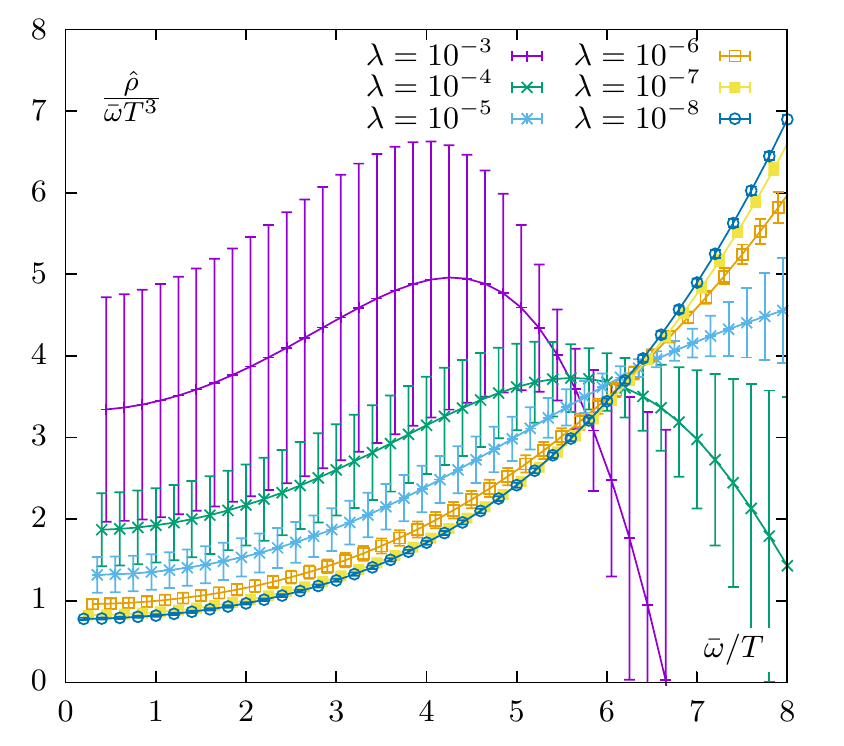}
}
\caption{The resolution function (left) and output spectral function (right) at $\bar{\omega}=0$ in shear channel at some selected $\lambda$ values from Backus-Gilbert analysis. 
}
\label{reso_spf_shear}
\end{figure*}

The technical difficulty in performing the spectral reconstruction can be traced in part to two issues, the finiteness of the number of data points and their noise. 
The first implies a discretization of the integral transform
\begin{align}
G(\tau)=\int_0^\infty \mathrm{d}\omega \rho(\omega) K(\tau,\omega) \rightsquigarrow G(\tau_i)=\sum_{i=1}^{N_\tau} \rho(\omega) \tilde{K}(\tau_i,\omega)
\end{align}
i.e. the underlying task is an inverse problem to find $\rho$ at a given $\omega$, schematically written as  $\rho= \sum_i\tilde{K}_i^{-1}G_i$. 

Consider an estimator $\hat\rho$ of the spectral function at a given $\bar\omega$ by (see e.g., \cite{Brandt:2015sxa,Brandt:2015aqk})
\begin{align}
\label{eq_spfestimator}
\hat\rho(\bar\omega)=f(\bar{\omega}) \int_0^\infty \mathrm{d}\omega  \,\delta(\bar\omega,\omega)\,\rho(\omega) \,f({\omega})^{-1}
\end{align}
where $f({\omega})$ is an arbitrary rescaling function and $\delta(\bar\omega,\omega)$ is a smooth function, normalized to $\int_0^\infty d\omega \delta(\bar\omega,\omega)=1$, that may be parametrized as $\delta(\bar \omega, \omega) = \sum_i q_i(\bar \omega) K(\tau_i, \omega)$ \cite{Backus1968}. 
This so-called resolution function acts as an averaging kernel that enables formulating the spectral function estimator as
\begin{align}
\label{eq_bgmweights}
    \hat{\rho}(\bar{\omega})=f(\bar{\omega})\sum_i q_i(\bar{\omega})\,G(\tau_i).
\end{align}
In this form it becomes clear that constructing $\hat\rho$, or by extension $\rho$, depends crucially on the number of coefficients available, i.e. the number of data points, and their behavior (how stable and regular the inverse is). Typically one is faced with a situation where the coefficients are large and highly fluctuating, requiring very precise determinations, but at the same time the connected matrix is nearly singular, requiring a regulator to be inverted safely. 
The added effect of noise in the data further complicates this situation as it affects the precision with which the coefficients can be determined. 

Keeping this in mind, one recipe to evaluate $\hat{\rho}(\bar{\omega})$ is given by the Backus-Gilbert method (BGM) \cite{Backus1968}. Construct the coefficients $q_i$ such that the width $\Gamma$, or spread, of the resolution function in $\omega$ becomes minimal, i.e., in the ideal case $\lim_{\Gamma\rightarrow 0}\hat\rho = \rho$. Then the solution can be shown to be:=
\begin{align}
q_i(\bar{\omega}) &= \frac { \sum_j W^{-1}_{ij} (\bar \omega) R(\tau_j) } { \sum_{kj} R(\tau_k) W^{-1}_{kj} (\bar \omega) R(\tau_j) }, 
\end{align}
\begin{align}
\begin{split}
W_{ij}(\bar \omega) &= \lambda\int_{0}^{\infty} \mathrm{d} \omega K(\tau_i,\omega) (\omega- \bar \omega)^2 K(\tau_j,\omega) \\ &\hspace{25ex}+ (1 - \lambda) S_{i j}, \\
R(\tau_i) &= \int_0^{\infty} \mathrm{d} \omega K(\tau_i,\omega)~~.
    \end{split}
\end{align}
Here we immediately introduced a regularization scheme $W_{ij} = \lambda W_{ij}^{no\,reg.} + (1-\lambda) S_{ij} $, where $S$ is the covariance matrix of the lattice correlators and $0\leq \lambda \leq 1$ is the regularization parameter. Other regularization schemes, such as the Tikhonov scheme where $S_{ij}=\mathds{1}$, have also been used in literature, see e.g. \cite{Astrakhantsev:2018oue}. Another recipe where the $q_i(\bar\omega)$ are determined with a fixed input resolution function was presented in \cite{Hansen:2019idp}.

In our implementation we further consider the rescaling function $f(\bar\omega)$ \cite{Brandt:2015aqk}. It rescales the spectral function inside the integral of Eq.~(\ref{eq_spfestimator}) prior to reconstruction and is or may be reintroduced afterwards. 
The coefficients $q_i$ are changed as a result and the procedure can be understood as related to a kernel transformation. In particular divergent behaviors of the kernel, such as that at $\omega\rightarrow 0$ can be handled in this way. Additionally certain well established, global trends of the spectral function can be built-in, for example the large frequency behavior $\sim \omega^4$. As such the procedure can also be seen as introducing prior information and some level of model dependence.
Here we consider the function $f=(\omega/T)^4/\tanh^3(\omega/4T)$ introduced to regularize the divergence at $\omega=0$ and to encode the information on the asymptotic trend. 

One key difficulty in the BGM, or any spectral reconstruction, is the determination of its errors, both statistical and systematic. The number of points, the rescaling function, the regularization parameter and the noise of the data all feed into the estimator result. 
Here, we focus on the impact of the regularization parameter $\lambda$. We also tested the impact of using different numbers of points and rescaling functions, but find that using the maximum number of points that have a stable solution and the above mentioned scaling function $f$ lead to the smallest spread of the resolution function. So in this study we use all the available data points. Note that $\lambda$ to some extent also controls the impact of noise given by the covariance through the regularization prescription.

Choosing $\lambda$ one would like to use the value which minimizes $\Gamma(\delta(\bar\omega,\omega))$ in the frequency window of interest.
In the left panel of Fig.~\ref{reso_spf_shear} we show the resolution function dependence for a broad range $\lambda$ in the shear channel. We see that the width is $\Gamma(\delta(\bar\omega,\omega))\sim 5 T$ except for the two smallest $\lambda$, which implies that the dependence of $\Gamma$ on $\lambda$ is weak.
At the same time, when plotting the obtained spectral functions depending on $\lambda$ in the right panel of Fig.~\ref{reso_spf_shear}, we see that the variance of the spectral function and crucially the value of the intercept at $\bar\omega=0$ depend strongly on this parameter. Based on the discussion above the increasing variance with $\lambda$ can be understood as insufficient regularization, while the decreasing variance with $\lambda$ but increasing width of the resolution implies the data and coefficients cannot be combined to form sharp, localized features.

Nevertheless, a robust result over a broad range in $\lambda$ implies a stable solution of the reconstruction. As such scanning through $\lambda$ in $(0,1)$ does suggest a lower bound for the intercept and thereby the viscosity. 
For the shear viscosity we find $\eta/T^3\geq 0.81$ (see right panel of Fig.~\ref{reso_spf_shear}). Similarly for bulk viscosity we obtain $\zeta/T^3\geq 0.059$. We can see the fit results determined in previous section safely lie in this range. 

One could imagine using a criterion for $\lambda$ based on the variance of the output spectral function instead of the spread of the resolution function, given the strong dependence observed,. The Morozov discrepancy principle \cite{Morozov1984} could be used for this: It states that $\delta\hat\rho(\bar\omega)/ \hat\rho(\bar\omega)=\overline{\delta G(\tau)}/G(\tau)$, where $\overline{\delta G(\tau)}$ denotes the average correlator variance. Since we are mainly interested in $\bar\omega=0$ one could impose this condition by matching $\delta\hat\rho(0)/ \hat\rho(0)=\delta G(T/2)/G(T/2)$, as the long-$\tau$ correlator data dominates the low-$\omega$ spectral function regime \cite{Aarts:2005hg}. This neglects the resolution function and the matching gives just a rough approximation to the more complicated underlying relation. However, applying this criterion we arrive at results for $\eta/T^3$ and $\zeta/T^3$ that agree with the quoted plateau values above.
% %%%%%%%%%%%%%%%%%%%%%%%%%%%%%%%%%%%%
% End of new version of BGM section
% %%%%%%%%%%%%%%%%%%%%%%%%%%%%%%%%%%%%

\section{Conclusion}
\label{sec:conclusion}

We have calculated the energy-momentum tensor correlators in both the shear and the bulk channel at $1.5T_c$ in the quenched approximation on five large and fine lattices.
To improve the signal-to-noise ratio we have applied both the gradient flow method and the blocking method.
We thoroughly studied the temperature corrections and the renormalization of the operators.
The correlators have been extrapolated first to the continuum limit and then to the $\tauf\rightarrow 0$ limit.
The final correlators are used to extract the shear and bulk viscosity based on perturbative models.
For the bulk channel, we find that the NLO spectral function can describe our lattice data when adding a transport part with appropriate interpolation.
For the shear channel we were unable to find a fit with better than $\chi^2$/d.o.f. = 2.
To further improve the fit quality, we need either a more flexible model or a better theoretical understanding of the expected spectral function.

\begin{table}[htb]       
\centering
\begin{tabular}{|c|c|c|c|}\hline
\diagbox[width=6em]{Model}{Measure}&
   $\zeta/T^3$ & $\eta/T^3$ \\ \hline
M1 & $\;$0.086(8)$\;$ & $\;$0.77(16)$\;$ \\ \hline
M2 & 0.133(10) & 1.09(15) \\ \hline
M3 & 0.303(31) & 2.46(54) \\ \hline
\end{tabular}
\caption{Bulk and shear viscosity fit results for three models, described in the previous section.  The errors are statistical only; the difference between different fit models represents a systematic error.  In each case, the NLO spectral function at large momentum was used in the fit.}
\label{tab:errors_fits}
\end{table}

In fitting our data, we find that the statistical errors are significantly smaller than the difference in fit values found from various fit Ans\"atz choices, despite relatively little difference in the fit quality from the different Ans\"atz choices.  This is summarized in Table~\ref{tab:errors_fits}.
Therefore we will estimate the lowest and highest value of viscosity to be the extreme values we found among the fit functions.
Using $s/T^3 = 5.098$ from \cite{Giusti:2016iqr}, our shear and bulk results become
%If we take the best estimates from our fits, include the statistical error, and estimate the systematical uncertainty using the difference between different fitting models, we can fix a window for shear viscosity and bulk viscosity %\todohaitao{needs updated numbers here and below}
\begin{equation}
    \begin{split}
        &\eta/s = 0.15 - 0.48,\ \ T=1.5T_c,\\
        &\zeta/s = 0.017 - 0.059 ,\ \ T=1.5T_c.
    \end{split}
    \label{shear_bulk_viscosity}
\end{equation}
%with $s/T^3=5.098$ taken from \cite{Giusti:2016iqr}. 
The lower estimates are above the lower bounds from the Backus-Gilbert analysis.
The upper bounds are based on a model which assumes that there \textsl{is} a relatively narrow feature near $\omega=0$, namely a Lorentzian-type peak with a width of $1 T$.
If a strongly-coupled medium does not support long-lived excitations, this assumption appears unlikely and the lower limit is more likely to be correct.
However, the data cannot definitively prove or disprove this theoretical prejudice.
The shear viscosity we obtained in Eq.~(\ref{shear_bulk_viscosity}) is close to the hydrodynamic estimate $1<(4\pi)\eta /s<2.5$ \cite{Song:2010mg}.

%Now we check whether our results agree with the relation $\zeta/\eta\approx 15(\frac{1}{3}-v_s^2)^2$  \cite{Weinberg:1971mx},
%or the AdS/CFT calculation at strong coupling $\zeta/\eta\propto (\frac{1}{3}-v_s^2)$ \cite{Buchel:2008uu}, 
% Guy objects to this because we can't check the functional dependence on $1/3 - v_s^2$ from one temperature, and this expected behavior seems to be an artifact of something about the AdS setup and how they break conformality, and is strongly expected NOT to hold for realistic QCD-like theories.
%where $v_s$ is the velocity of sound in the medium.
%In the latter equation the authors proposed that $m_f=0$ and $\mathcal{N}=2^*$ can more accurately reflect the fact that RHIC produces QGP close to its criticality. In this case at $1.5T_c$ $\zeta/\eta\approx 0.17-0.61$. 
%If we calculate $v_s$ using \cite{Meyer:2007dy}
%\begin{align}
%    1-3v_s^2=\frac{4}{3}\frac{\epsilon-3P}{\epsilon+P}[1+\mathcal{O%}(\alpha_s)]
%\end{align}
%up to LO and the values of $\epsilon$ and $P$ from \cite{Giusti:2016iqr}, we find the former equation gives $\zeta/\eta=0.259$, lying in the range from latter equation. Our results suggest a range $0.078 \leq \zeta/\eta \leq 0.173$, favoring the latter equation.

In our opinion, there are two pressing tasks to further improve on this work.
The first is to find better models for the spectral function's behavior at low to intermediate frequencies $\omega \sim [1-5]T$.
This will allow a fitting extraction which makes maximal use of the high-quality data which is now available.
The second task is to extend these results to the unquenched case.
This is not just a matter of performing much more expensive unquenched simulations.
It is also necessary to understand the renormalization of the more-complicated unquenched stress tensor operator at the percent level, which appears to be possible but quite challenging.
Some progress in this direction has been made recently by Dalla Brida $et$ $al$ \cite{DallaBrida:2020gux}, but precision studies including gradient flow do not yet exist.
We leave these developments for future work.

All data from our calculations, presented in the figures of this paper, can be found in \cite{datapublication}.
%Morozov discrepancy principle \cite{book}
%bulk 0.73(0.07) 
%shear 0.92(0.05)
%\cite{Dudal:2019gvn}

\section*{Acknowledgements}
All authors acknowledge support by the Deutsche For\-schungs\-ge\-mein\-schaft
(DFG, German Research Foundation) through the CRC-TR 211 'Strong-interaction matter under extreme conditions'– Project No. 315477589 – TRR 211. A.F. acknowledges support by the Ministry of Science and Technology Taiwan (MOST) under Grant No. 111-2112-M-A49-018-MY2.
The computations in this work were performed on the GPU cluster at Bielefeld University using \texttt{SIMULATeQCD} suite \cite{Mazur:2023lvn,Altenkort:2021fqk,mazur2021}.
We thank the Bielefeld HPC.NRW team for their support.

\section*{Appendix}
\appendix

\section{Uncertainties of the renormalization constants}
\label{sec:error_renorm}

In Sec.~\ref{sec_renormalization} we introduce the renormalization coefficients $c_1$, $c_2$ to be used with the traceless and pure-trace stress tensor operators respectively.
The coefficient $c_2$ is determined very accurately from an analytical perturbative series, so there is no need to specify it further than through Eq.~(\ref{coeff_UE2}).

The coefficient $c_1$ depends more strongly on flow depth and lattice spacing, and our nonperturbative determination contains statistical error bars.
Therefore, we present tabulated values with errors in Table \ref{tab:errors_c1} for future reference.
The errors are an important ingredient in our error analysis and error budget, though the errors in the correlation functions themselves are typically larger.

\begin{table}[htb]       
\centering
\begin{tabular}{|c|c|c|c|c|c|}\hline
\diagbox[width=5.2em]{$\vphantom{\Big|}\!\tau_FT^2\;$}{$\;N_s^{3\vphantom{|}}{\times} N_t\!$}&
  $64^3\times 16$ & $80^3\times 20$ & $96^3\times 24$ & $120^3{\times} 30$ & $144^3{\times} 36$ \\ \hline
0.00158 & 5.40(6) & 5.23(1) & 5.22(1) & 5.24(2) & 5.21(2) \\ \hline
0.00203 & 5.14(5) & 5.00(1) & 5.01(1) & 5.03(2) & 5.01(2) \\ \hline
0.00254 & 4.93(4) & 4.82(1) & 4.83(1) & 4.85(1) & 4.85(2) \\ \hline
0.00310 & 4.75(3) & 4.67(1) & 4.68(2) & 4.69(2) & 4.69(2) \\ \hline
0.00372 & 4.59(2) & 4.52(1) & 4.53(2) & 4.54(2) & 4.55(2) \\ \hline
0.00439 & 4.45(2) & 4.38(1) & 4.39(2) & 4.40(2) & 4.41(2) \\ \hline
0.00513 & 4.32(2) & 4.25(1) & 4.27(1) & 4.28(2) & 4.29(2) \\ \hline
0.00591 & 4.20(1) & 4.14(1) & 4.15(1) & 4.17(2) & 4.17(2) \\ \hline
0.00861 & 3.88(1) & 3.83(1) & 3.84(1) & 3.87(1) & 3.87(1) \\ \hline
\end{tabular}
\caption{$c_1\times 10$ at selected flow times in the valid flow time window for all the lattices.  The value, and errors, at flow times between the listed values are reliably determined by interpolation.}
\label{tab:errors_c1}
\end{table}

\section{Uncertainties in the temperature correction}
\label{sec:error-T-correction}

\begin{figure}[ht]
\centerline{\includegraphics[width=0.5\textwidth]{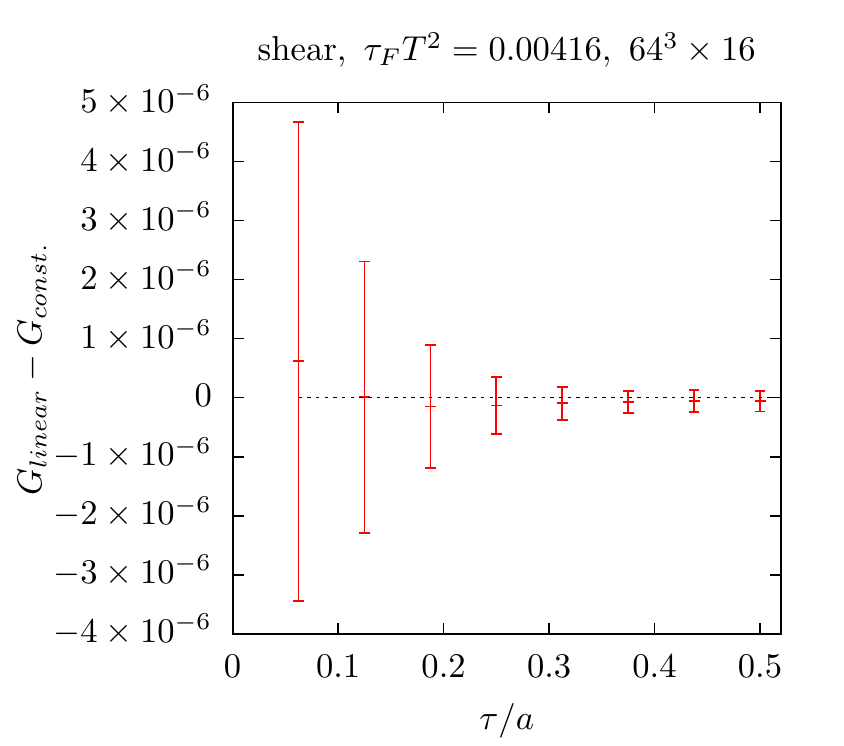}}
\caption{The difference of correlators obtained in two different ways of treating the slope of the correlators with respect to $\tau T$ in the temperature correction. Note only data at $\tau T > 0.35$ can be used in later flow-time extrapolation according to the flow-time limitation.}
\label{fig_linear_const}
\end{figure}

In correcting for the slight temperature variation between our lattices, we made the assumption that the temperature dependence in the spectral function is approximately separation-independent.
Looking at Fig.~\ref{G1G2}, it also looks reasonable to assume that the correction is linear in $\tau T$.
Therefore, we consider this Ansatz, and consider the difference between the two assumptions as a source of systematic uncertainty.
This difference is shown for the specific case of the shear channel, the $64\times 16$ lattice, and the flow depth $\tau_FT^2=0.00416$ (same as the one used in Fig.~\ref{cont-extrap} and Fig.~\ref{fig_linear_const}).
Note that at this flow time the usable data points must have $\tau T > 0.35$.
The figure shows that the difference in these approaches generates an effect which is small compared to, e.g., statistical errors.

\section{The uncertainties in the double extrapolation}
\label{sec:error-double-extrap}

The errors in the continuum extrapolation, shown in Fig.~\ref{cont-extrap}, are statistical errors arising from the data and from $c_1$.
For the data presented, the bulk-viscous extrapolations are almost flat, but this is not true in general when we consider other flow depths.
We tested for the need for a linear term in the extrapolation by repeating the fits assuming no lattice spacing dependence (simply averaging data across all lattices).
This increases the $\chi^2/$d.o.f. (averaged over all flow times valid for the flow-time extrapolation) from 1.78 to 4.75, showing that linear extrapolation is in fact needed.

\begin{figure}[ht]
\centerline{\includegraphics[width=0.5\textwidth]{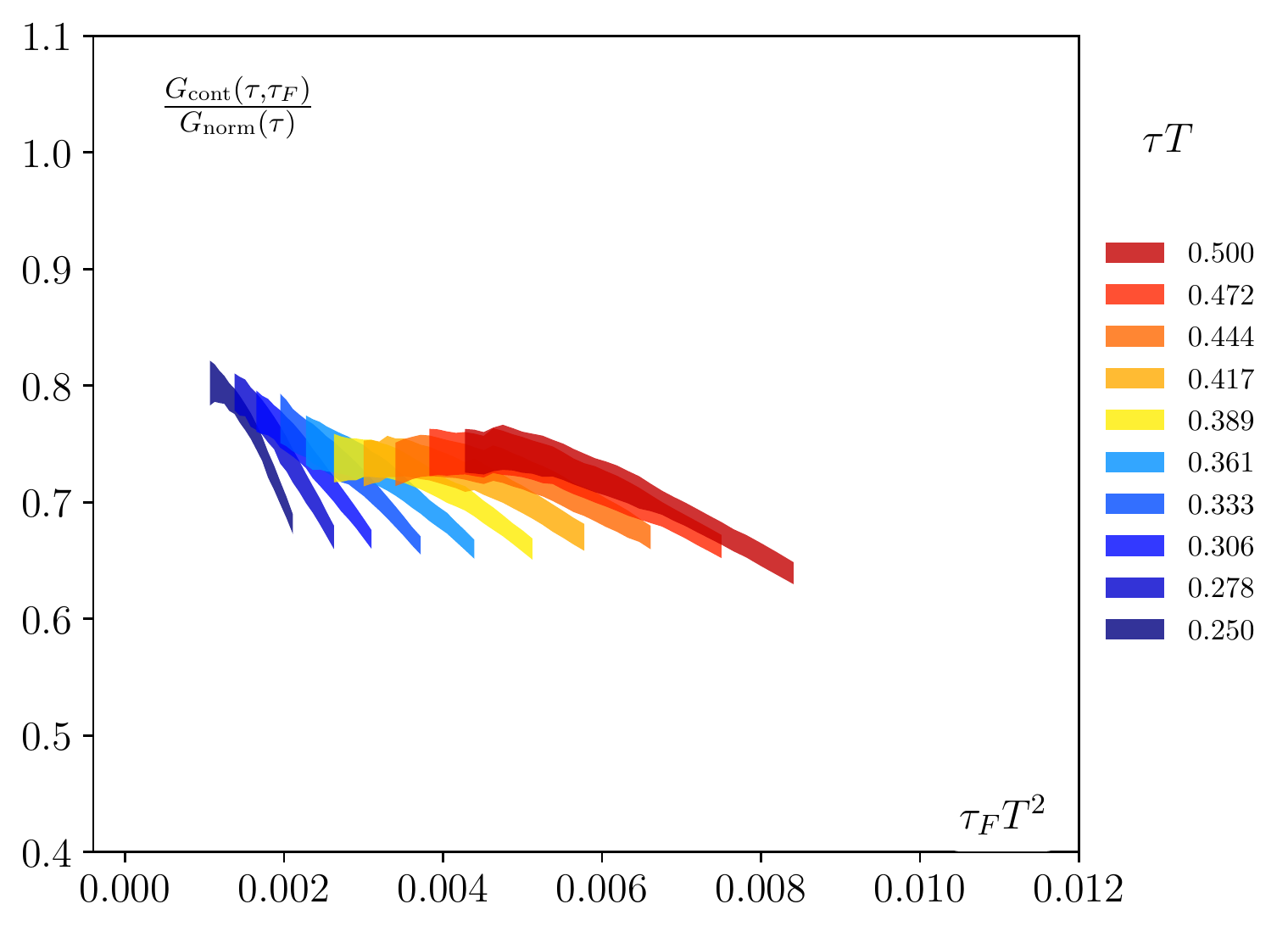}
}
\centerline{\includegraphics[width=0.5\textwidth]{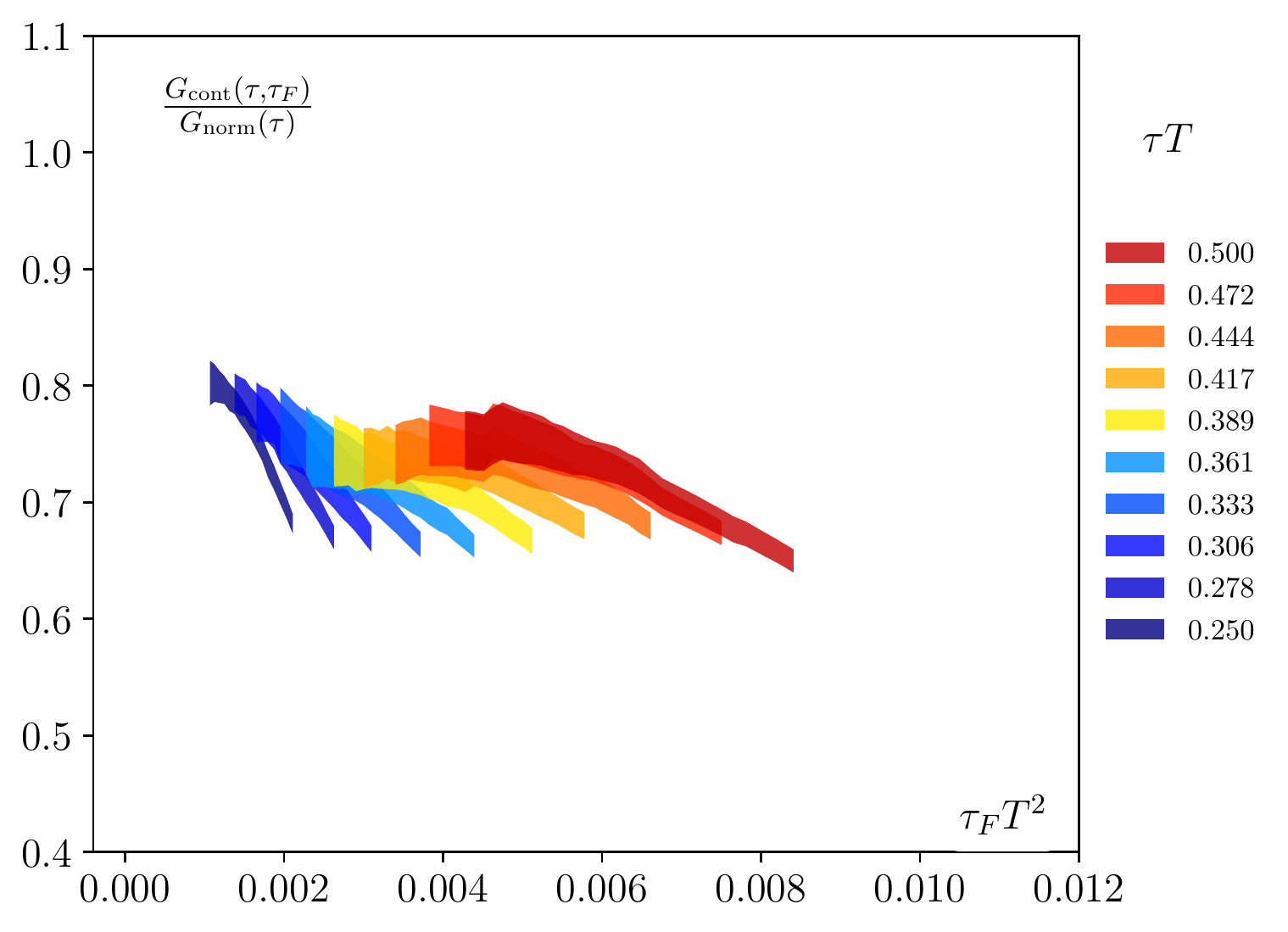}}
\caption{The difference of continuum-extrapolated correlators in shear channel with (top) and without (bottom) the coarsest lattice.}
\label{fig_4lattice}
\end{figure}

We also tried continuum extrapolation excluding the coarsest lattice $64^3\times 16$, which suffers the most severe discretization effects.
We compare the continuum extrapolated correlators in Fig.~\ref{fig_4lattice}, taking the shear channel as an example. We can see the central values only change very mildly, while the errors increase slightly, as expected.
Such changes will affect our estimate of the viscosities by less than the quoted statistical errors.

% Similar questions can be asked about the extrapolation shown in Fig8. How does the extrapolation and error change if one less or one more term is used in the fit?

Next consider the extrapolation to zero flow depth.
In the main text we argue that the operator product expansion predicts flow-depth effects which are polynomial in $(\tauf/\tau^2)$, at least where this parameter is small.
We can then compare three small-flow fit models:  a constant, a linear extrapolation, and a quadratic fit:
\begin{equation}
    \begin{split}
        &\mathrm{F1:}\ G(\tauf/\tau^2) = A\\
        &\mathrm{F2:}\ G(\tauf/\tau^2) = A + B \tauf/\tau^2\\
        &\mathrm{F3:}\ G(\tauf/\tau^2) = A + B \tauf/\tau^2+ C \tauf^2/\tau^4
    \end{split}
\end{equation}

In Table~\ref{tab:errors_flow_model} we summarize the flow extrapolated results in the shear channel for the relative error of $A$ in percentage (averaged over $\tau T\in [0.22, 0.5]$) and averaged $\chi^2$/d.o.f.\ using each of these models, all performed in the same flow-time windows which we use in the main text.
\begin{table}[htb]       
\centering
\begin{tabular}{|l|c|c|c|}\hline
\diagbox[width=6em]{Measure}{Model}&
   F1 & F2 & F3 \\ \hline
$\delta A [\%]$ & 0.40 & 1.84 & 12.5 \\ \hline
$\chi^2$/d.o.f. & 14.4 & 2.14 & 1.84 \\ \hline
\end{tabular}
\caption{Flow-extrapolated results using different kinds of models.}
\label{tab:errors_flow_model}
\end{table}
It can be seen that fitting the data in Fig.\ref{tauF-extrap} without a linear term leads to a very poor fit, with $\chi^2/$d.o.f. values in the range of [5.5, 28.4]. Adding a $\tau_F^2$ term over-fits the data, dramatically increasing the errors, but is not justified by the very small improvement in $\chi^2$.

\section{Relative importance of statistical error sources}
\label{xxx}

Statistical errors arise both in our determined $c_1,c_2$ values (normalization coefficients for the stress tensor) and directly as statistical fluctuations in the measured correlators.
To compare the relative importance of these two sources, we have repeated our analysis but leaving out the errors in $c_1$ (the errors in $c_2$ are so small that they make no difference).
Table \ref{tab:errors_sources_shear} shows that leaving out the errors in $c_1$ (middle column) only slightly reduces the final statistical error in the fully extrapolated correlation function.
Therefore, the errors in the Euclidean data are, in practice, dominated by statistical errors in the determined correlation functions.

\begin{table}[htb]       
\centering
\begin{tabular}{|c|c|c|}\hline
\diagbox[width=6em]{$\tau T$}{\%-error}&
   case 1 & case 2\\ \hline
0.222 & 2.30 & 2.67 \\ \hline
0.250 & 1.41 & 1.58 \\ \hline
0.278 & 1.73 & 1.87 \\ \hline
0.306 & 1.48 & 1.77 \\ \hline
0.333 & 1.56 & 1.84 \\ \hline
0.361 & 1.64 & 1.86 \\ \hline
0.389 & 1.50 & 1.70 \\ \hline
0.417 & 1.75 & 1.87 \\ \hline
0.444 & 1.69 & 1.78 \\ \hline
0.472 & 1.69 & 1.75 \\ \hline
0.500 & 1.51 & 1.57 \\ \hline
\end{tabular}
\caption{The (un)importance of the statistical error in $c_1$.
The first column lists separations.
The third column is the percent error in the determined fully extrapolated correlation function.
The middle column is the error we would find if we neglect the error in $c_1$. }
\label{tab:errors_sources_shear}
\end{table}

\iffalse
\begin{table}[htb]       
\centering
\begin{tabular}{|l|c|c|c|}\hline
\diagbox[width=6em]{$\tau T$}{Measure}&
   case 1 & case 3\\ \hline
0.222 & 3.25 & 2.51 \\ \hline
0.250 & 3.26 & 2.61\\ \hline
0.278 & 2.93 & 2.55\\ \hline
0.306 & 2.84 & 2.40\\ \hline
0.333 & 2.10 & 1.89\\ \hline
0.361 & 1.81 & 1.69\\ \hline
0.389 & 1.63 & 1.51\\ \hline
0.417 & 1.82 & 1.68\\ \hline
0.444 & 2.08 & 1.86\\ \hline
0.472 & 3.02 & 2.58\\ \hline
0.500 & 2.60 & 2.23\\ \hline
\end{tabular}
\caption{$\chi^2$/d.o.f. for the flow-time-to-zero extrapolation corresponding to case 1 and case 3 in the shear channel.}
\label{tab:chisq_shear}
\end{table}
\fi

\bibliographystyle{apsrev4-1}
\bibliography{Bibliography}

\end{document}